\documentclass[]{mn2e}
\usepackage{epsfig}
\usepackage{graphicx}
\usepackage{mathtools}
\usepackage{pgf}
\def\gsim{\vcenter{\hbox{$>$}\offinterlineskip\hbox{$\sim$}}}
\def\lsim{\vcenter{\hbox{$<$}\offinterlineskip\hbox{$\sim$}}}
\title[Star formation history in NGC\,147 and NGC\,185]
{Long period variable stars in NGC\,147 and NGC\,185. I. Their star formation
histories}
\author[R.H.\ Golshan et al.]{
Roya H.\ Golshan$^{1,2}$,
Atefeh Javadi$^1$,
Jacco Th.\ van Loon$^3$,
Habib Khosroshahi$^1$
\newauthor
and Elham Saremi$^{1,4}$\\
$^1$School of Astronomy, Institute for Research in Fundamental Sciences (IPM),
    Tehran, 19395-5531, Iran\\
$^2$Department of Physics, Isfahan University of Technology, Isfahan,
    84156-83111, Iran\\
$^3$Lennard-Jones Laboratories, Keele University, ST5 5BG, UK\\
$^4$Physics Department, University of Birjand, Birjand 97175-615, Iran}
\pagerange{\pageref{firstpage}--\pageref{lastpage}}
\pubyear{2016}
\begin{document}
\maketitle
\label{firstpage}
\begin{abstract}
NGC\,147 and NGC\,185 are two of the most massive satellites of the Andromeda
galaxy (M\,31). Close together in the sky, of similar mass and morphological
type dE, they possess different amounts of interstellar gas and tidal
distortion. The question therefore is, how do their histories compare? Here we
present the first reconstruction of the star formation histories of NGC\,147
and NGC\,185 using long-period variable stars. These represent the final phase
of evolution of low- and intermediate-mass stars at the asymptotic giant
branch, when their luminosity is related to their birth mass. Combining
near-infrared photometry with stellar evolution models, we construct the mass
function and hence the star formation history. For NGC\,185 we found that the
main epoch of star formation occurred 8.3 Gyr ago, followed by a much lower,
but relatively constant star formation rate. In the case of NGC\,147, the star
formation rate peaked only 7 Gyr ago, staying intense until $\sim3$ Gyr ago,
but no star formation has occurred for at least 300 Myr. Despite their similar
masses, NGC\,147 has evolved more slowly than NGC\,185 initially, but more
dramatically in more recent times. This is corroborated by the strong tidal
distortions of NGC\,147 and the presence of gas in the centre of NGC\,185.
\end{abstract}
\begin{keywords}
stars: AGB and post-AGB --
stars: luminosity function, mass function --
stars: oscillations --
galaxies: evolution --
galaxies: individual: NGC\,147, NGC\,185 --
galaxies: stellar content
\end{keywords}
%=========================================================================== 1
\section{Introduction}

Dwarf galaxies are the most abundant type of galaxies in the universe,
spanning a huge range in stellar mass, luminosity and surface brightness. They
come in two main flavours: dwarf Irregulars (dIrrs) and dwarf
Spheroidal/Ellipticals (dSph/dEs). These are distinguished principally by the
significant interstellar medium (ISM) in dIrrs and almost complete lack
thereof in dSphs/dEs, but their star formation histories also differ, with
dIrrs being ``younger'' (Battinelli \& Demers 2006). Most dSph/dEs are located
in the dense inner regions of galaxy groups and clusters, close to massive
spiral or elliptical galaxies. On the other hand, dIrrs are often more
isolated. This strongly suggests that environmental effects transform dIrrs
into dSph/dEs.

The Star Formation History (SFH) is one of the most important tracers of the
evolution of galaxies. We have developed a novel method to use Long-Period
Variable stars (LPVs) to reconstruct the SFH, which we have successfully
applied the Local Group galaxies M\,33 (Javadi, van Loon \& Mirtorabi 2011;
Javadi et al.\ 2016) and the Magellanic Clouds (Rezaeikh et al.\ 2014). The
most evolved stars with low to intermediate mass, at the tip of the Asymptotic
Giant Branch (AGB) show brightness variations on timescales of $\approx100$ to
$>1000$ days due to radial pulsation. LPVs represent the most luminous phase
in their evolution, $\sim3000$--60,000 L$_\odot$, and reach their maximum
brightness at near-infrared wavelengths. Intermediate-mass AGB stars may
become carbon stars as a result of the dredge up of carbon synthesized in the
helium thermal pulses; the resulting change in opacity reddens their colours.
Since the maximum luminosity attained on the AGB relates to the star's birth
mass, we can use the brightness distribution function of LPVs to construct the
birth mass function and hence the Star Formation Rate (SFR) as a function of
time. In this paper we present the results from application of this method to
the Local Group dEs NGC\,147 and NGC\,185.

% FIGURE 1
\begin{figure*}
\centerline{\hbox{
\epsfig{figure=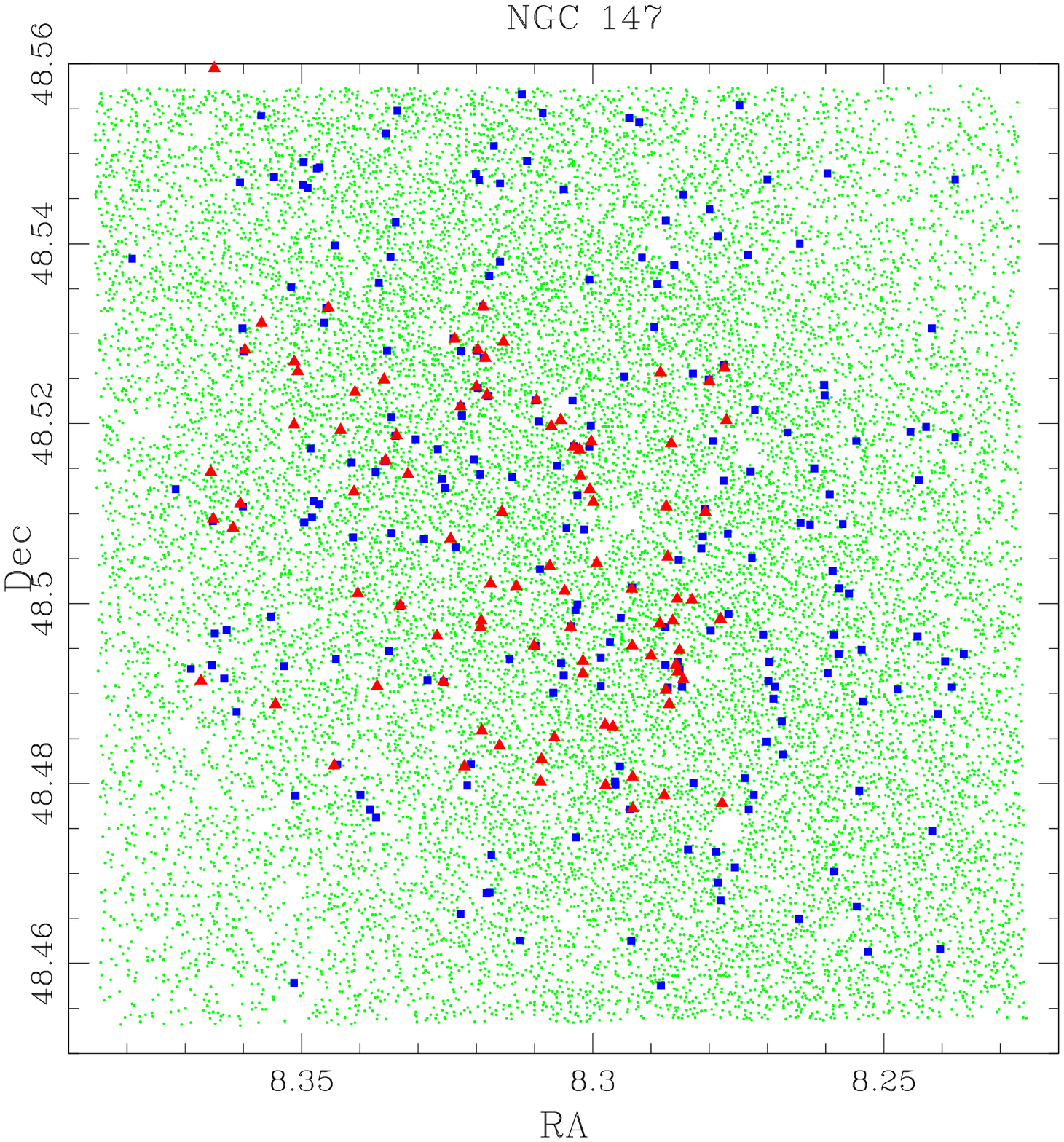,width=88mm}
\epsfig{figure=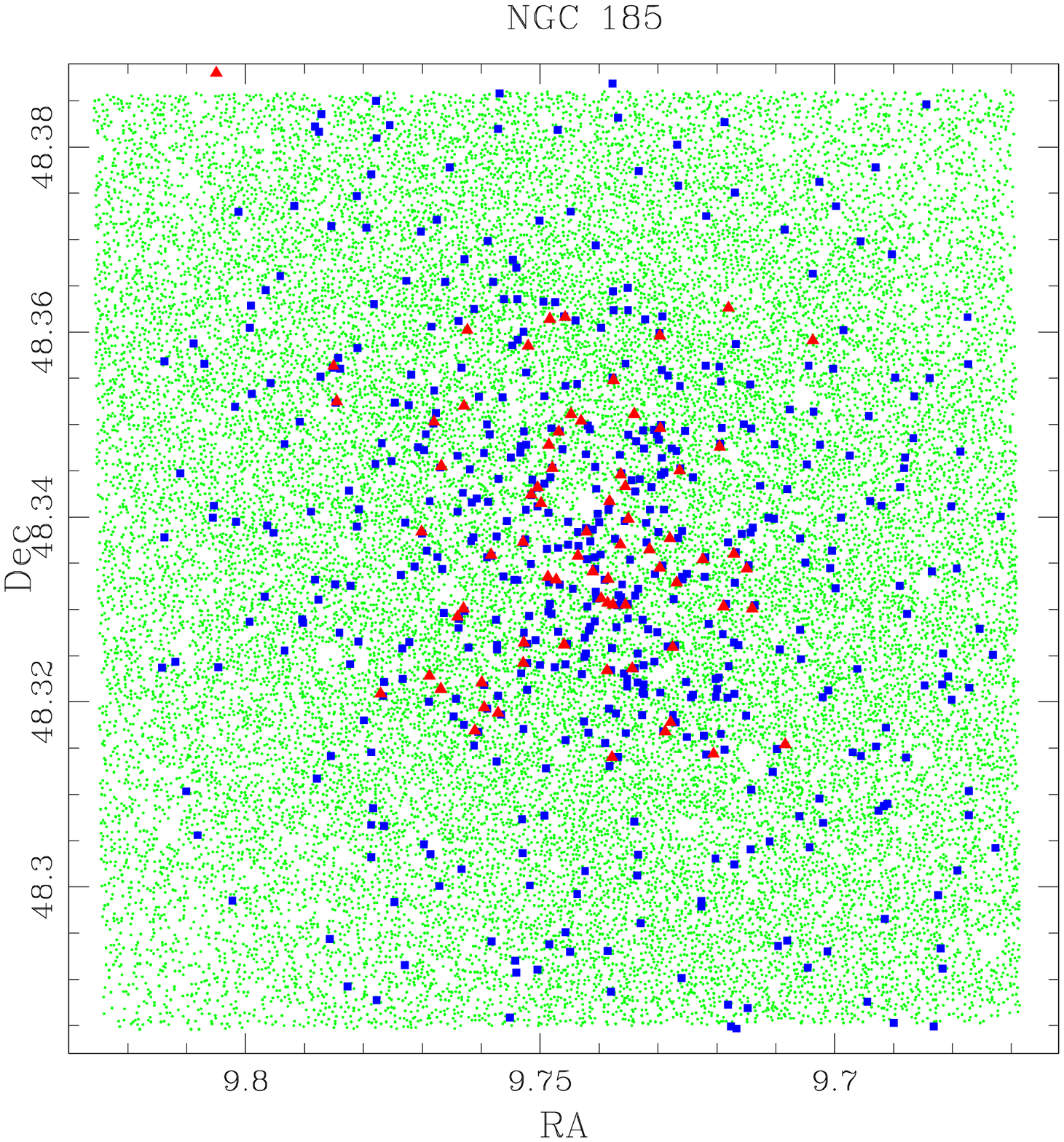,width=88mm}
}}
\caption{Spatial distribution of AGB stars (green dots; Nowotny et al.\ 2003),
LPVs (blue squares; Lorenz et al.\ 2011) and carbon stars (red triangles;
Sohn et al.\ 2006; Kang et al.\ 2005) across NGC\,147 ({\it left}) and
NGC\,185 ({\it right}).}
\label{fig:fig1}
\end{figure*}

NGC\,147 and NGC\,185 were selected for the availability of suitable data of a
significant number of LPVs (Lorenz et al.\ 2011), but also in their own right.
Both are satellites of Andromeda (M\,31), the most massive member of the Local
Group. NGC\,147 and NGC\,185 share fundamental properties such as luminosity
($M_{\rm V}\sim 16$; Crnojevi\'c et al.\ 2014) and velocity dispersion (25 km
s$^{-1}$; Geha et al.\ 2010), but they differ in many other aspects. NGC\,185
contains some gas, dust and shows evidence for recent star formation in its
central regions, while NGC\,147 is destitute of gas and dust and shows no sign
of recent star formation activity (Young \& Lo 1997; Welch, Sage \& Mitchell
1998; Marleau, Noriega-Crespo \& Misselt 2010; De Looze et al.\ 2016). The
mean metallicity of NGC\,147 based on photometry of the red giant branch is
[M/H]$_{\rm NGC\,147}\approx-1.1$ to $-1$, which is slightly higher than that of
NGC\,185, [M/H]$_{\rm NGC\,185}\approx-1.3$ to $-1.1$ (Davidge 1994; McConnachie
et al.\ 2005; Geha et al.\ 2010). Spectroscopic measurements indicate the
presence of a more metal-rich population, with [M/H]$_{\rm NGC\,147}=-0.5\pm0.1$
and [M/H]$_{\rm NGC\,185}=-0.9\pm0.1$ (Vargas, Geha \& Tollerud 2014). The
latter displays a metallicity gradient (Vargas et al.\ 2014; Crnojevi\'c et
al.\ 2014), reaching [M/H]$_{\rm NGC\,185}\sim-2$ at the edge (Ho et al.\ 2015).
NGC\,147 and NGC\,185 are very close in projection on the sky ($58^{\prime}$),
leading some to argue that they may be a bound galaxy pair (van den Bergh
1998). However, kinematic evidence suggests that they may not be
gravitationally bound (Geha et al.\ 2010; Watkins, Evans \& van de Ven 2013).
NGC\,147 is tidally distorted, whereas NGC\,185 is not (Ferguson \& Mackey
2016). Their distances have been estimated (McConnachie et al.\ 2005) at
$d_{\rm NGC 147}=675\pm27$ kpc (distance modulus $\mu_{\rm NGC 147}=24.15$ mag)
and $d_{\rm NGC 185}=616\pm26$ kpc ($\mu_{\rm NGC 185}=23.95$ mag).

In this paper we construct the birth mass function of LPVs and derive the SFH
in the inner $6\rlap{.}^\prime5\times6\rlap{.}^\prime5$ regions of NGC\,147 and
NGC\,185. In section \ref{sec:sec2} we introduce the photometric catalogues.
Section \ref{sec:sec3} explains our method. In section \ref{sec:sec4} we
present the results. Finally, a discussion and conclusions follow in sections
\ref{sec:sec5} and \ref{sec:sec6}. In an Appendix, we provide additional
results for different assumptions regarding the metallicity.

%=========================================================================== 2
\section{Data}
\label{sec:sec2}

We take advantage of a number of published photometric catalogues (see below).
The dimensions of NGC\,147 and NGC\,185 are
$13\rlap{.}^\prime2\times7\rlap{.}^\prime8$ and
$11\rlap{.}^\prime7\times10\rlap{.}^\prime0$
\footnote{http://ned.ipac.caltech.edu/}, however the data only cover the
central $6\rlap{.}^\prime5\times6\rlap{.}^\prime5$ regions (see Fig.\
\ref{fig:fig1} -- one thing that is immediately apparent is that the AGB stars
in NGC\,147 are much less centrally concentrated than in NGC\,185).

%------------------------------------------------------------------------- 2.1
\subsection{The optical data}

As part of a photometric survey of Local Group galaxies to resolve AGB stars,
Nowotny et al.\ (2003) obtained images of NGC\,147 and NGC\,185 in four
optical filters, viz.\ the broad-band visual $V$ and near-infrared Gunn $i$
and filters centred on the TiO and CN molecular bands, respectively, using the
Nordic Optical Telescope (NOT) in the year 2000. They used the ALFOSC
instrument, with a pixel size of $0\rlap{.}^{\prime\prime}189$ pixel$^{-1}$
yielding a $6\rlap{.}^\prime5\times6\rlap{.}^\prime5$ field of view (FoV). 
Although this will include the majority of the stars, our survey is not totally complete in terms of area. Figure
\ref{fig:fig1} shows the observed areas, corresponding to physical sizes of
$1.27\times1.27$ kpc$^2$ on NGC\,147 and $1.16\times1.16$ kpc$^2$ on NGC\,185.
A total of 18\,300 and 26\,496 AGB stars were detected within these regions in
NGC\,147 and NGC\,185, respectively (green dots in figure \ref{fig:fig1}).

%------------------------------------------------------------------------- 2.2
\subsection{The near-infrared data}
%....................................................................... 2.2.1
\subsubsection{LPVs}

% FIGURE 2
\begin{figure*}
\centerline{\hbox{
\epsfig{figure=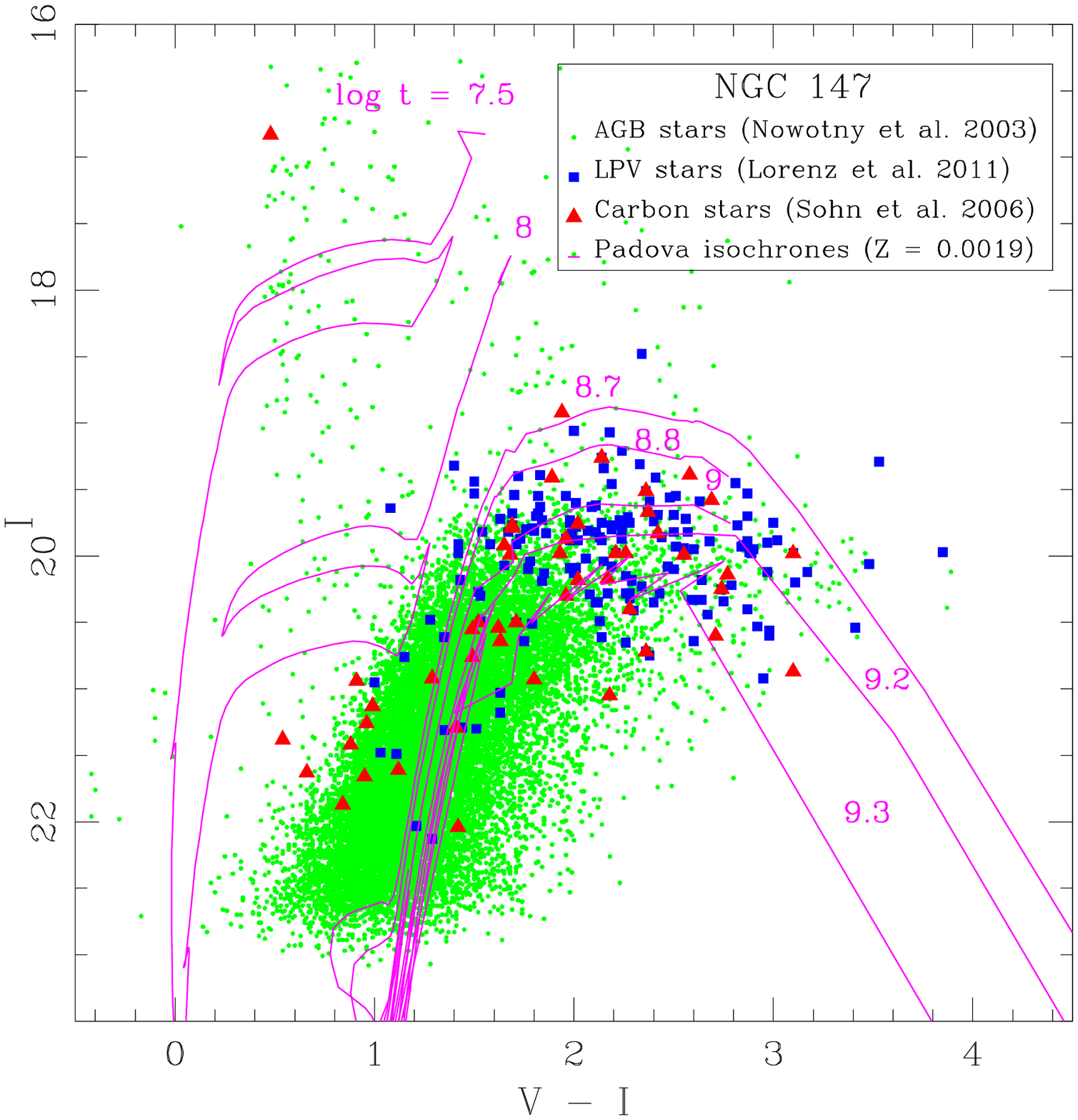,width=58mm}
\epsfig{figure=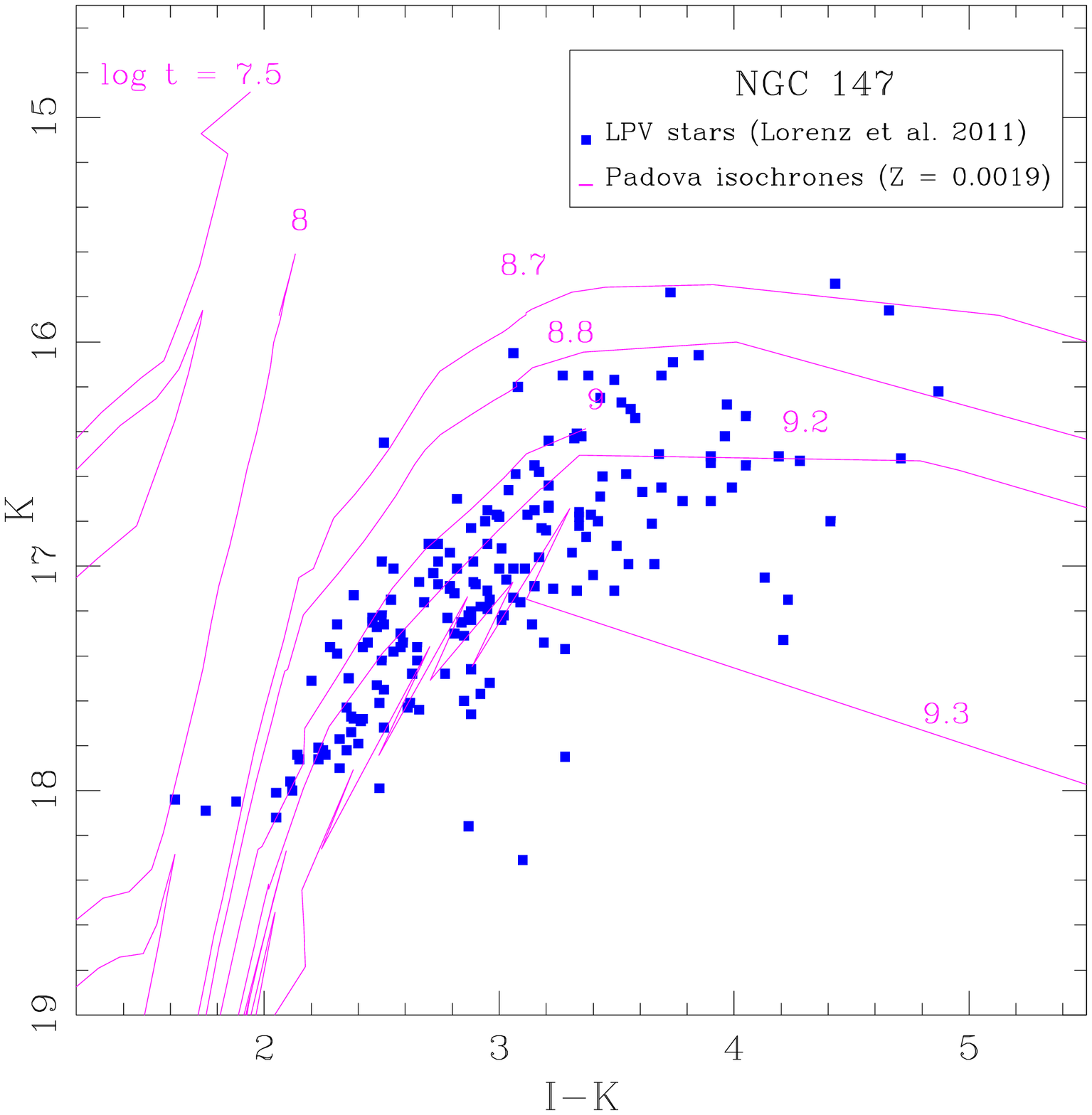,width=58mm}
\epsfig{figure=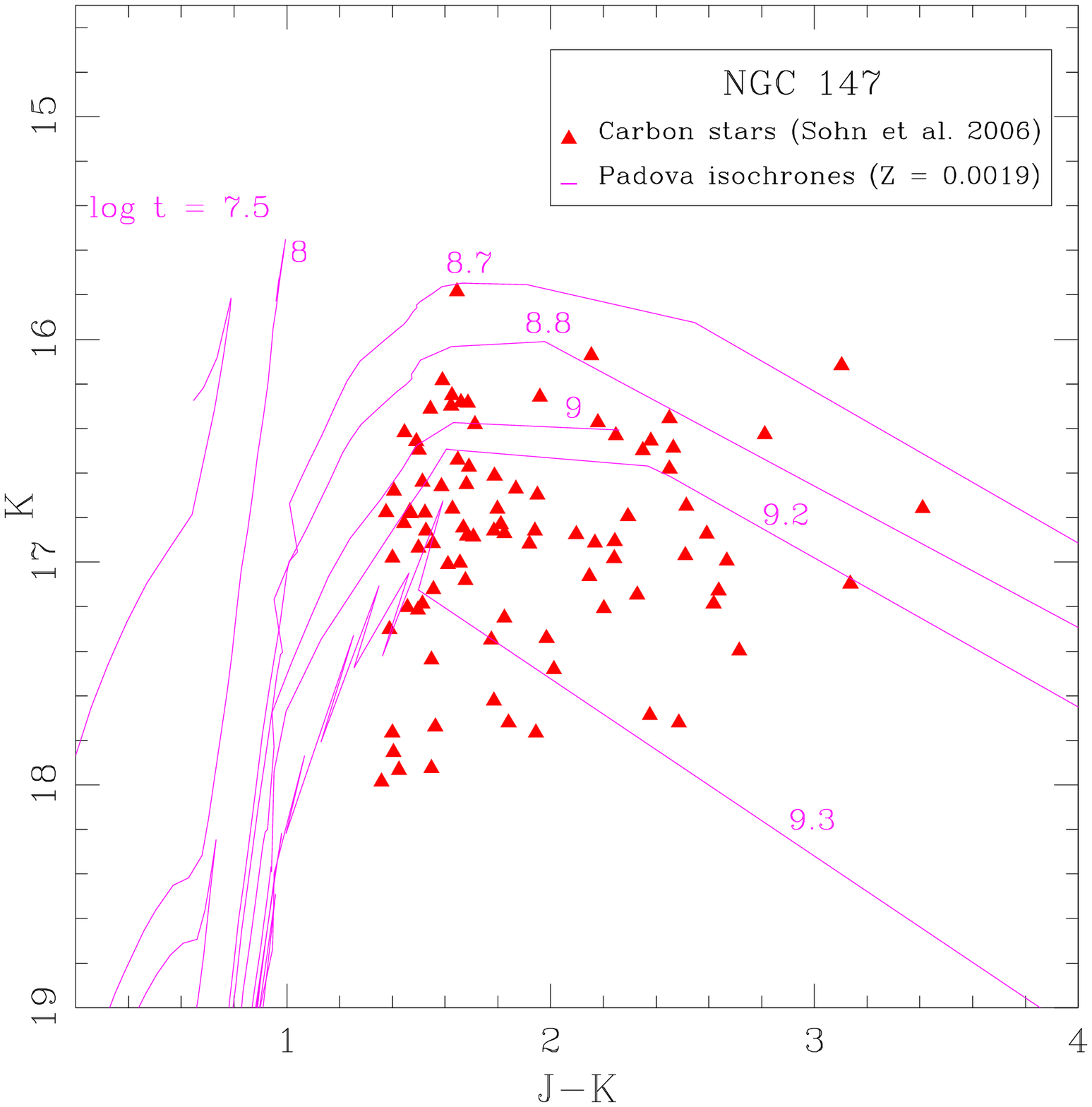,width=58mm}
}}
\caption{CMDs of NGC\,147. Isochrones from Marigo et al.\ (2008) are labelled
with logarithmic ages.}
\label{fig:fig2}
\end{figure*}

% FIGURE 3
\begin{figure*}
\centerline{\hbox{
\epsfig{figure=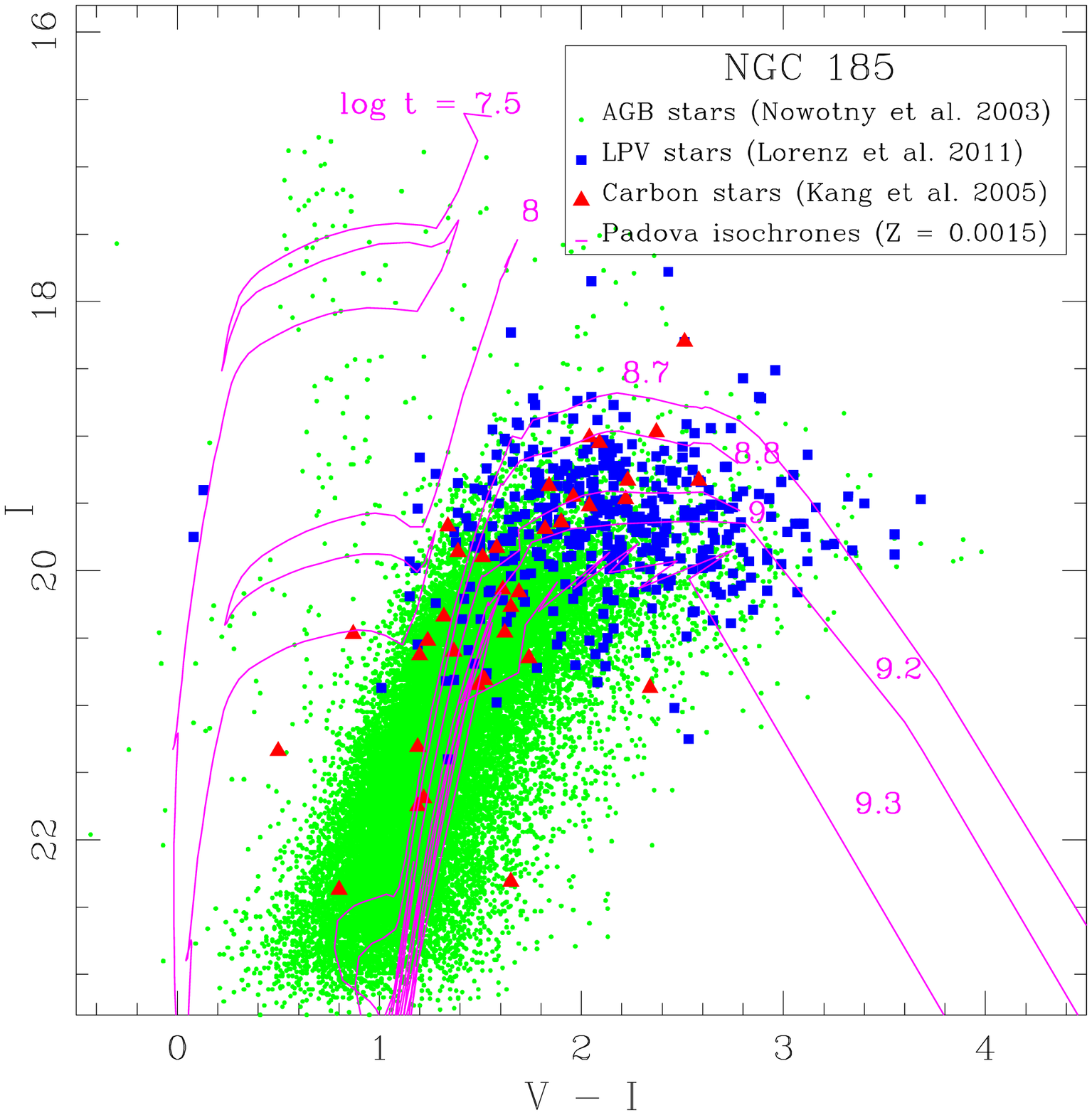,width=58mm}
\epsfig{figure=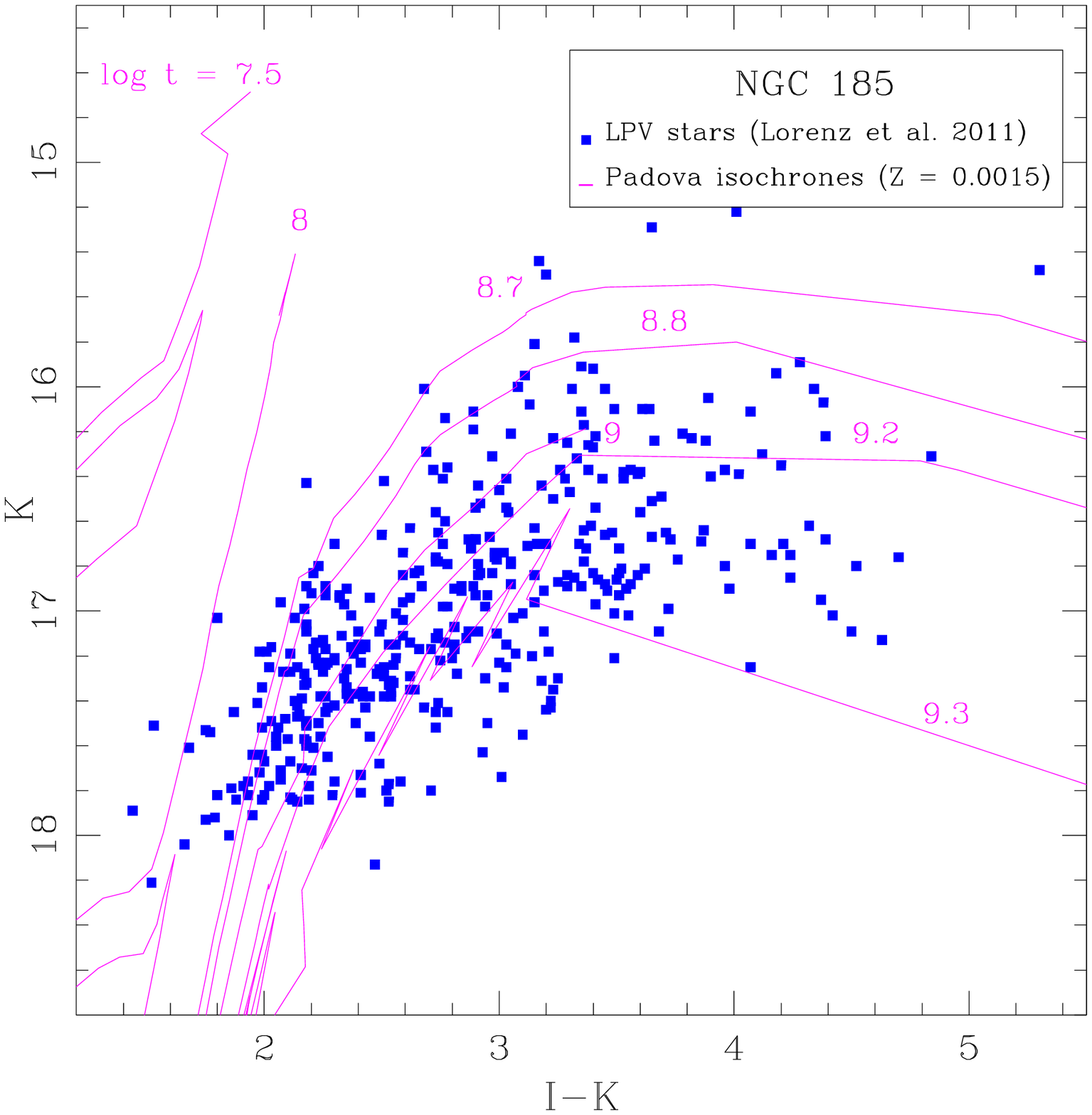,width=58mm}
\epsfig{figure=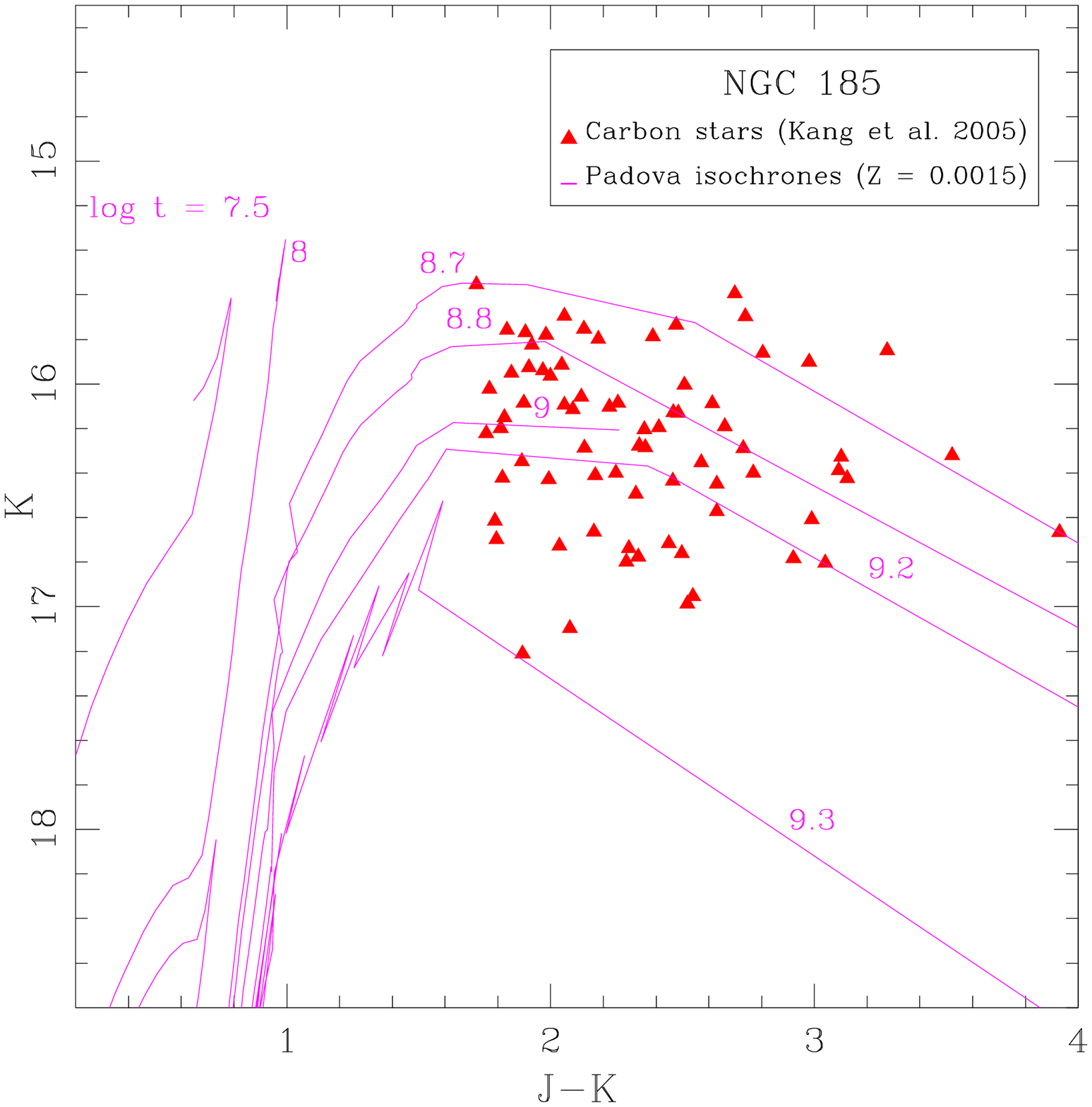,width=58mm}
}}
\caption{CMDs of NGC\,185. Isochrones from Marigo et al.\ (2008) are labelled
with logarithmic ages.}
\label{fig:fig3}
\end{figure*}

For our method to derive the SFH, we need to identify the LPVs among the AGB
stars. For this we have used the catalogue of LPVs from Lorenz et al.\ (2011).
Like Nowotny et al.\ (2003), they observed NGC\,147 and NGC\,185 with ALFOSC
on the NOT, using the Gunn $i$ band, on 38 nights between October 2003 and
February 2006. They also observed NGC\,147 and NGC\,185 once in the $K$ band,
using NOTCam (on the NOT) in September 2004. NOTcam yields a field-of-view of
$4^\prime\times4^\prime$; to cover the ALFOSC field-of-view they created a mosaic
of four overlapping images with NOTcam. The LPVs were identified by means of
the image subtraction technique. The resulting catalogue contains 213 LPVs in
the direction of NGC\,147 and 513 LPVs in the direction of NGC\,185. The loci
of the LPVs are depicted in figure \ref{fig:fig1} with blue squares.
As can be seen in figure \ref{fig:fig1}, the central region of NGC 185 is denser towards the centre. Hence, the identification of variable stars in the central regions is incomplete because of crowding.

%....................................................................... 2.2.2
\subsubsection{Carbon stars}

We have also made use of the catalogues of carbon stars published in Sohn et
al.\ (2006) and Kang et al.\ (2005). They obtained near-infrared images with
the Canada--France--Hawai'i Telescope in June 2004, using the CFHTIR imager
and $J$, $H$ and $K^\prime$ filters, each field covering
$3\rlap{.}^\prime6\times3\rlap{.}^\prime6$. Sohn et al.\ (2006) present the
photometry of 91 carbon stars in NGC\,147, and Kang et al.\ (2005) present the
photometry of 73 carbon stars in NGC\,185. The loci of the carbon stars are
depicted in figure \ref{fig:fig1} with red triangles.

%------------------------------------------------------------------------- 2.3
\subsection{Colour--magnitude diagrams}

The $K$-band photometry is the best suited for calculating the SFR, because
the spectral energy distributions of stars near the tip of the AGB peak at
near-infrared wavelengths, minimising uncertainties in bolometric corrections,
and interstellar and circumstellar extinction is much less important than at
optical wavelengths. Because carbon stars can be used to trace (part of) the
AGB, we have cross matched the LPV and carbon star catalogues, using the
average of $K$-band magnitudes for stars in common. We can thus examine the
location of the AGB stars in a colour--magnitude diagram (CMD), in comparison
with isochrones from the stellar evolution models from the Padova group
(Marigo et al.\ 2008) that we use later in our technique to derive the SFH.
The CMDs of NGC\,47 and NGC\,185 are presented in figures \ref{fig:fig2} and
\ref{fig:fig3}, respectively.

The optical CMDs (left panels of figures \ref{fig:fig2} and \ref{fig:fig3})
clearly show the densely populated AGB, from $(V-I)\sim1$ to $\sim2$ mag. The
smaller number of brighter/bluer stars are likely Galactic foreground dwarfs.
The carbon stars occupy the bright/red portion of the AGB as these are in
advanced stages of AGB evolution; likewise, the LPVs occupy the region where
the AGB terminates, thus validating our premise that LPVs can be used to
determine the luminosity at the tip of the AGB.

Padova isochrones (Marigo et al.\ 2008) are consistent with the photometry
with the adopted distance moduli and metallicities of $Z=0.0019$ for NGC\,147
and $Z=0.0015$ for NGC\,185. They include the effects of circumstellar dust
formed in the winds from the most evolved AGB stars, at $(V-I)\,\gsim\,2.5$
mag.

Our method, which is explained in section \ref{sec:sec3}, Javadi et al.\
(2011, 2016) and Rezaeikh et al.\ (2014), is based on the luminosity peak of
LPVs in the $K$ band. Figures \ref{fig:fig2} and \ref{fig:fig3} show the
location of the LPVs in the near-infrared CMDs (middle panels), as well as the
carbon stars for comparison (right panels). Clearly, the vast majority of LPVs
are several Gyr old ($\log t>9$), with the youngest $\sim400$--500 Myr old
($\log t=8.6$--8.7).

%=========================================================================== 3
\section{Method: star formation history}
\label{sec:sec3}

The SFH of a galaxy is a measure of the rate at which gas mass was converted
into stars over a time interval in the past. Or in other words, it is the SFR,
$\xi$ (in M$_\odot$ yr$^{-1}$), as a function of time. The amount of stellar
mass, $dM$, created during a time interval, $dt$, is:
\begin{equation}
dM(t) = \xi(t)\ dt.
\label{eq:eq1}
\end{equation}
The number of formed stars are related to the total mass by the following
equation:
\begin{equation}
dN(t) = \frac{\int_{\rm min}^{\rm max}f_{\rm IMF}(m)\ dm}
{\int_{\rm min}^{\rm max}f_{\rm IMF}(m)m\ dm}\ dM(t),
\label{eq:eq2}
\end{equation} 
where $f_{\rm IMF}$ is the Initial Mass Function (IMF). We use the IMF defined
in Kroupa (2001):
\begin{equation}
f_{\rm IMF} = Am^{-\alpha},
\label{eq:eq3}
\end{equation}
where $A$ is the normalization constant and $\alpha$ is defined depending on
the mass range:
\begin{equation}
\alpha = \left\{\begin{array}{cclcl}
+0.3\pm0.7 & {\rm for} & {\rm min} & < \frac{m}{{\rm M}_\odot} < & 0.08 \\
+1.3\pm0.5 & {\rm for} & 0.08      & < \frac{m}{{\rm M}_\odot} < & 0.5 \\
+2.3\pm0.3 & {\rm for} & 0.5       & < \frac{m}{{\rm M}_\odot} < & {\rm max} \\
\end{array}\right.
\label{eq:eq4}
\end{equation}
The minimum and maximum mass were assumed to be 0.02 and 200 M$_\odot$,
respectively.

We need to relate this to the number of stars, $N$, which are variable at the
present time. If stars with mass between $m(t)$ and $m(t+dt)$ are LPVs at the
present time, then the number of LPVs created between times $t$ and $t+dt$ is
\begin{equation}
dn(t) = \frac{\int_{m(t)}^{m(t+dt)}f_{\rm IMF}(m)\ dm}
{\int_{\rm min}^{\rm max}f_{\rm IMF}(m)\ dm}\ dN(t).
\label{eq:eq5}
\end{equation}
Substituting equation \ref{eq:eq1} and \ref{eq:eq2} in equation \ref{eq:eq5}
gives
\begin{equation}
dn(t) = \frac{\int_{m(t)}^{m(t+dt)}f_{\rm IMF}(m)\ dm}
{\int_{\rm min}^{\rm max}f_{\rm IMF}(m)m\ dm}\ \xi(t)\ dt.
\label{eq:eq6}
\end{equation}

We are considering an age bin of $dt$, to determine $\xi(t)$. The number of
LPVs observed in this age bin, $dn^\prime$, depends on the duration of the
evolutionary stage during which the variability occurs:
\begin{equation}
dn^\prime(t) = \frac{\delta t}{dt}\ dn(t).
\label{eq:eq7}
\end{equation}

Finally, by combining the above equations we obtain a relation to calculate
the SFR based on LPV counts:
\begin{equation}
\xi(t) = \frac{\int_{\rm min}^{\rm max}f_{\rm IMF}(m)m\ dm}
{\int_{m(t)}^{m(t+dt)}f_{\rm IMF}(m)\ dm}\ \frac{dn^\prime(t)}{\delta t}.
\label{eq:eq8}
\end{equation}

To obtain the SFR we need to determine the individual stars' masses, ages
($t$) and durations of pulsation ($\delta t$). For this we rely on theoretical
models. The most appropriate theoretical models for our purpose are the Padova
models (Marigo et al.\ 2008), for the following reasons:\\
$\bullet$ They have prepared models for the required range in birth mass
($0.8<M<30$ M$_\odot$) by combining models for intermediate-mass stars
($M<7$ M$_\odot$) with those for more massive stars ($M>7$ M$_\odot$; Bertelli
et al.\ 1994)) and corrections for low-mass, low-metallicity AGB tracks
(Girardi et al.\ 2010);\\
$\bullet$ They are based on computations carried out through the entire
thermal pulsing AGB until they enter the post-AGB phase, in a manner that is
consistent with the computation of the preceding evolutionary stages. The
models account for the third dredge-up mixing of the stellar envelope as a
result of the helium-burning pulses, and the increased luminosity of massive
AGB stars undergoing hot bottom burning (Iben \& Renzini 1983);\\
$\bullet$ The models account for the molecular opacity in the cool atmospheres
of red giants, including changes as a result of the transformation from
oxygen-dominated M-type AGB stars to carbon stars in the approximate
birth-mass range of $M\sim1.5$--4 M$_\odot$ (Girardi \& Marigo 2007);\\
$\bullet$ They include predictions for dust production in the winds of LPVs
and the associated attenuation and reddening of the synthetic photometry;\\
$\bullet$ They include predictions for the radial pulsation;\\
$\bullet$ They have been carefully transformed to various optical and infrared
photometric systems;\\
$\bullet$ They are available via a user-friendly internet interface.\\
One shortcoming is that these models do not account for the final part of the
evolution of super-AGB stars, i.e.\ stars in the mass range
$0.7<\log M/{\rm M}_\odot<1.2$--1.3.

We note that the PARSEC+COLIBRI models are the latest Padova models and we use the Padova models that incorporate the relevant physics already. However, according to Marigo et al. (2013), the difference
between the effective temperatures predicted by the two sets of models is
negligible, and the same is true for the temperature at the base of the
convective envelope. In the same paper they also show that their new models
reproduce perfectly the core-mass luminosity relation adopted in the previous models, as well as confirming the predictions from alternative models (from Karakas et al. 2002). Taken together, this means that the observational properties do not change markedly between these models, and that the relationship between empirical quantities and theoretical parameters remains essentially unchanged and robust.

\begin{figure*}
\centerline{\hbox{
\epsfig{figure=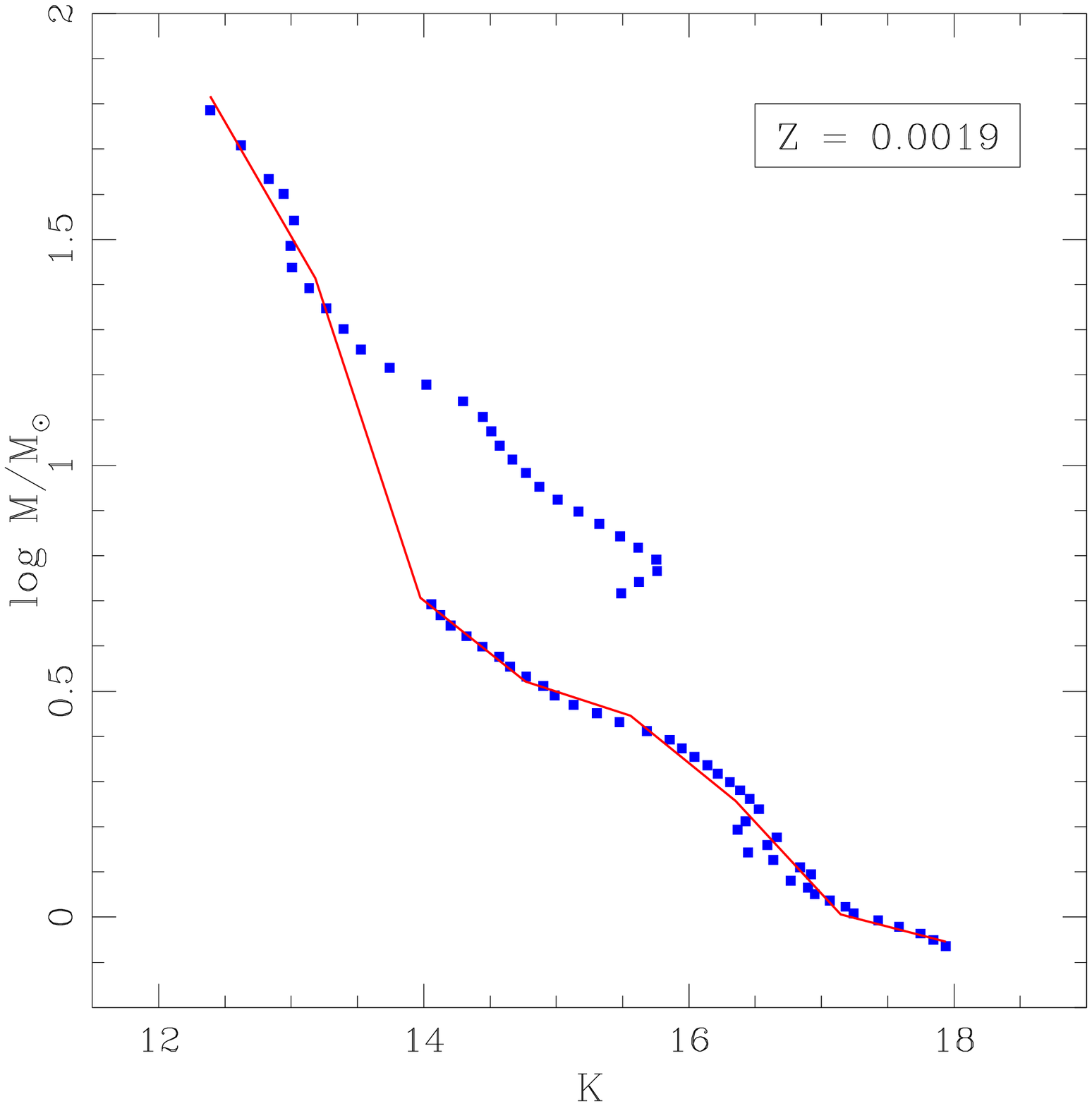,width=58mm}
\epsfig{figure=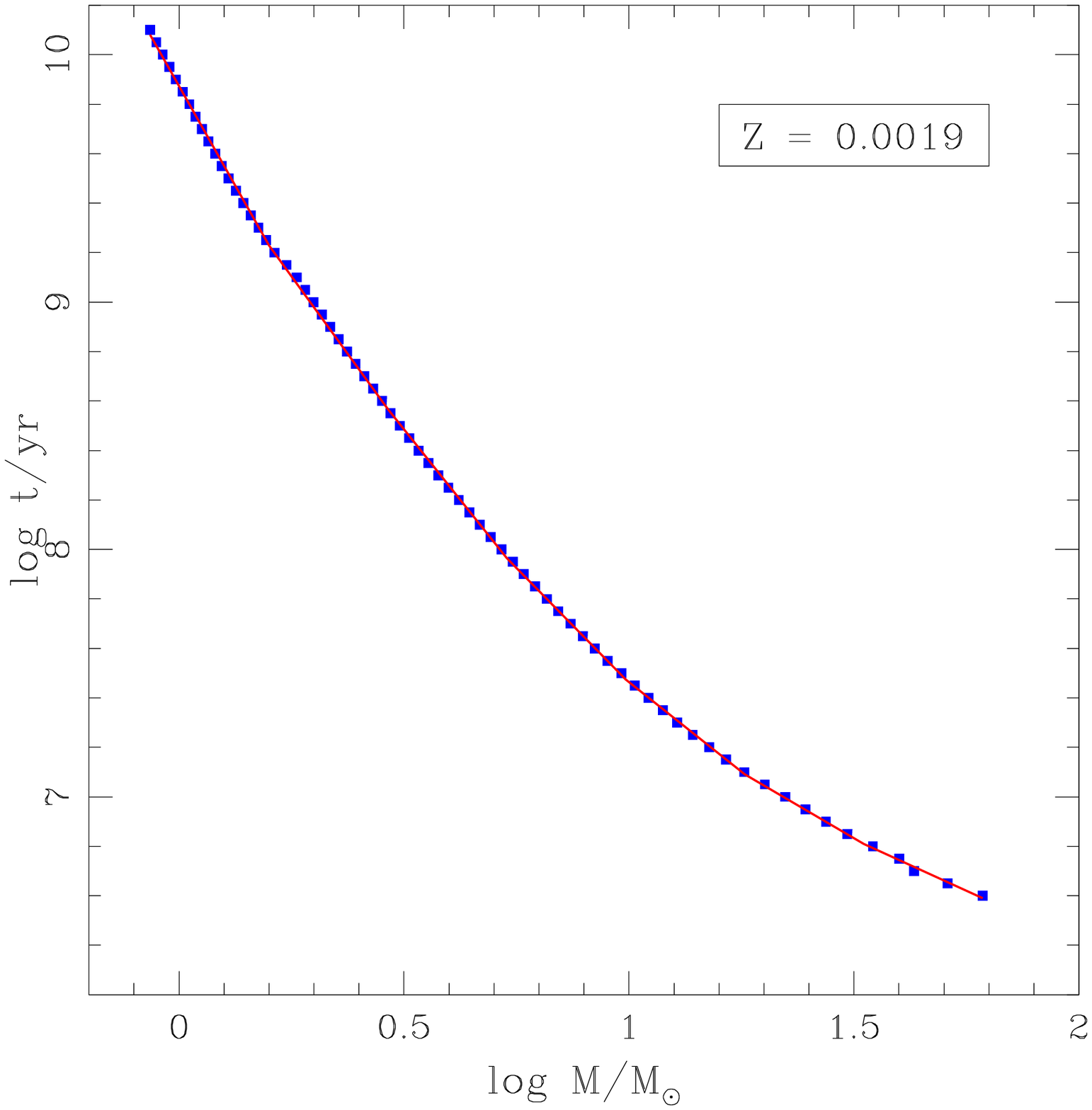,width=58mm}
\epsfig{figure=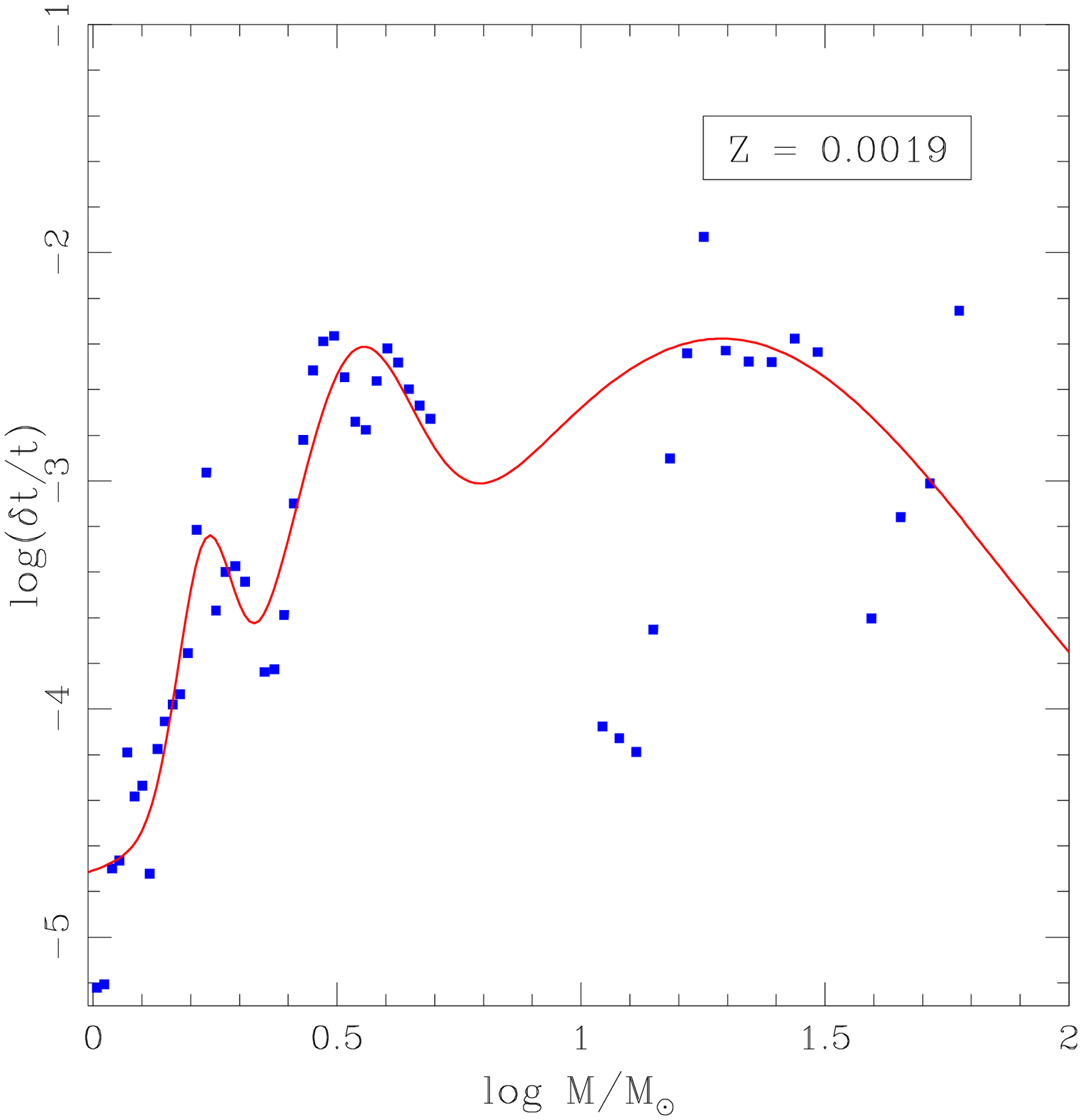,width=58mm}
}}
\caption[]{The dots refer to models from Marigo et al. (2008); ({\it Left:}) Mass--luminosity relation (in the $K$ band) for
$Z=0.0019$ and a distance modulus of $\mu_{\rm NGC 147}=24.15$ mag. The solid
lines are linear spline fits, in which the function is interpolated over the
super-AGB phase to massive red supergiants. ({\it Middle:}) Same, for the
mass--age relation. ({\it Right:}) Same, for the mass--pulsation duration
relation. The points show the ratio of pulsation duration to age, versus mass;
the solid lines are multiple-Gaussian fits, interpolated over the super-AGB
regime.}
\label{fig:fig4}
\end{figure*}

LPVs have reached the maximum luminosity on the AGB. On that premise, we have
constructed a mass--$K$-band magnitude relation for the metallicity and
distance modulus of NGC\,147 ($Z=0.0019$ and $\mu=24.15$ mag; left panel of
figure \ref{fig:fig4}). All the figures and tables related to NGC\,185 and to
other metallicities used in this paper are presented in the Appendix. 

There is
an obvious excursion towards fainter $K$-band magnitudes for super-AGB stars,
because their evolution to brighter $K$-band magnitudes has been omitted from
the models. We thus interpolate over that range in mass (see Javadi et al.\
2011); however, since we do not have any super-AGB stars in any of these dwarfs, we effectively only use the part of the Mass-Luminosity (ML) relation that is modelled carefully. In other words, the brightest stars have $K>15$ mag so we are in the regime of stars for which the models give reliable predictions. The coefficients of the piecewise linear relations between $K$-band
magnitude and mass are listed in Table \ref{tab:tab1}.

% FIGURE 4

% TABLE 1
\begin{table}
\caption[]{Relation between birth mass and $K$-band magnitude, $\log M=aK+b$,
for a distance modulus of $\mu_{\rm NGC 147}=24.15$ mag.}
\begin{tabular}{ccr}
\hline\hline
\multicolumn{3}{c}{$Z=0.0019$} \\
\hline
$a$              & $b$                     & validity range       \\
\hline
$-0.508\pm0.052$ &         $8.116\pm0.676$ &        $K\leq13.182$ \\
$-0.893\pm0.048$ & \llap{1}$3.185\pm0.649$ & $13.182<K\leq13.975$ \\
$-0.234\pm0.048$ &         $3.974\pm0.688$ & $13.975<K\leq14.767$ \\
$-0.095\pm0.044$ &         $1.931\pm0.673$ & $14.767<K\leq15.560$ \\
$-0.238\pm0.038$ &         $4.151\pm0.606$ & $15.560<K\leq16.352$ \\
$-0.317\pm0.032$ &         $5.437\pm0.534$ & $16.352<K\leq17.145$ \\
$-0.076\pm0.042$ &         $1.311\pm0.471$ &        $K>17.145$    \\
\hline
\end{tabular}
\label{tab:tab1}
\end{table}

We apply the following extinction corrections to dusty LPVs according to the
isochrones in the middle panels of figures \ref{fig:fig2} and \ref{fig:fig3}:
\begin{equation}
K _{\rm cor} = K+a_{\rm LPV}[(I-K)-(I-K)_0],
\label{eq:eq9}
\end{equation}
and to carbon stars according to the isochrones in the right panels of figures
\ref{fig:fig2} and \ref{fig:fig3}:
\begin{equation}
K _{\rm cor} = K+a_{\rm carbon}[(J-K)-(J-K)_0],
\label{eq:eq10}
\end{equation}
where $a_{\rm LPV}=0.24$ and $a_{\rm carbon}=0.59$ are the average slopes of the
tracks for these two types of stars in their respective CMDs, and
$(I-K)_0=2.6$ mag and $(J-K)_0=1.4$ mag are their mean colours at the point at
which their tracks bend towards lower luminosity and redder colour.

The mass--age relation is shown in the middle panel of figure \ref{fig:fig4}.
This relates the birth mass of an LPV observed at the present time, to the
time elapsed since its birth. The coefficients of this relation are listed in
table \ref{tab:tab2}, for the metallicity adopted for NGC\,147.

% TABLE 2
\begin{table}
\caption[]{Relation between age and birth mass, $\log t=a\log M+b$.}
\begin{tabular}{ccr}
\hline\hline
\multicolumn{3}{c}{$Z=0.0019$} \\
\hline
$a$              & $b$             & validity range           \\
\hline
$-3.221\pm0.030$ & $9.874\pm0.002$ &       $\log{M}\leq0.200$ \\
$-2.511\pm0.027$ & $9.732\pm0.009$ & $0.200<\log{M}\leq0.465$ \\
$-2.278\pm0.029$ & $9.624\pm0.018$ & $0.465<\log{M}\leq0.729$ \\
$-1.856\pm0.032$ & $9.317\pm0.028$ & $0.729<\log{M}\leq0.993$ \\
$-1.442\pm0.037$ & $8.906\pm0.041$ & $0.993<\log{M}\leq1.257$ \\
$-1.062\pm0.043$ & $8.428\pm0.059$ & $1.257<\log{M}\leq1.522$ \\
$-0.833\pm0.053$ & $8.078\pm0.087$ &       $\log{M}>1.522$    \\
\hline
\end{tabular}
\label{tab:tab2}
\end{table}

Massive stars spend less time in the LPV phase than lower mass stars, but a
larger fraction of their entire lifetime (right panel of figure
\ref{fig:fig4}). The theoretical models were parameterised by a set of three
Gaussian functions (table \ref{tab:tab3}, for the metallicity adopted for
NGC\,147).

% TABLE 3
\begin{table}
\caption[]{Relation between the relative pulsation duration and birth mass,
$\log(\delta t/t)=D+
\Sigma_{i=1}^3a _i\exp\left[-(\log M [{\rm M} _\odot]-b_i)^2/2c_i^2\right]$.}
\begin{tabular}{ccccc}
\hline\hline
\multicolumn{5}{c}{$Z=0.0019$} \\
\hline
$D$    & $i$ & $a$  & $b$  &  $c$  \\
\hline
$-4.9$ &  1  & 1.02 & 0.23 & 0.056 \\
       &  2  & 1.43 & 0.53 & 0.129 \\
       &  3  & 2.52 & 1.29 & 0.576 \\
\hline
\end{tabular}
\label{tab:tab3}
\end{table}

To obtain the SFH, the following procedure was applied to each individual
star:\\
$\bullet$ De-redden the $K$-band magnitude;\\
$\bullet$ Use table \ref{tab:tab1} to determine the birth mass;\\
$\bullet$ Use table \ref{tab:tab2} to determine the age of the star;\\
$\bullet$ Use table \ref{tab:tab3} to calculate the pulsation duration;\\
$\bullet$ As the number of young, massive LPVs is much smaller than the number of old, low-mass LPVs, to obtain uniform uncertainties in the SFR values we choose the age bin size such that they have equal numbers of stars. The error on the SFR in each bin is calculated using the following relation:
$errorSFR[i] = \sqrt{n[i]} \times SFR[i]/n[i]$
where n is the number of stars in each bin i.

%=========================================================================== 4
\section{Results: star formation histories}
\label{sec:sec4}

Using the photometric catalogues of LPVs and carbon stars (section
\ref{sec:sec2}) and applying the method explained in detail in section
\ref{sec:sec3}, we estimate the SFRs in NGC\,147 and NGC\,185 during the broad
time interval from 300 Myr to 13 Gyr ago.

%------------------------------------------------------------------------- 4.1
\subsection{Assuming constant metallicity over time}
\label{subsec:subsec4.1}

\begin{figure*}
\centerline{\hbox{
\epsfig{figure=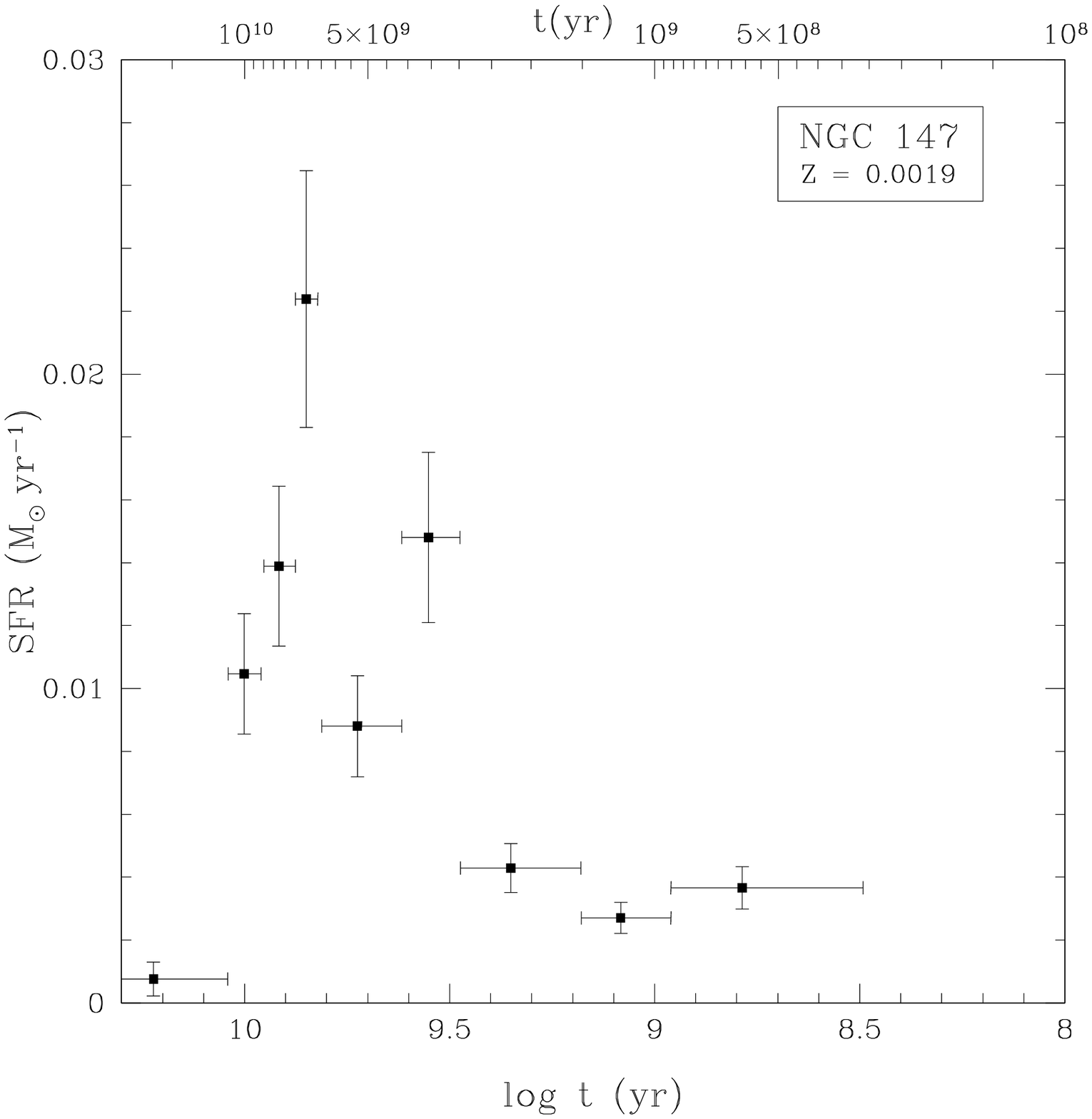,width=88mm}
\epsfig{figure=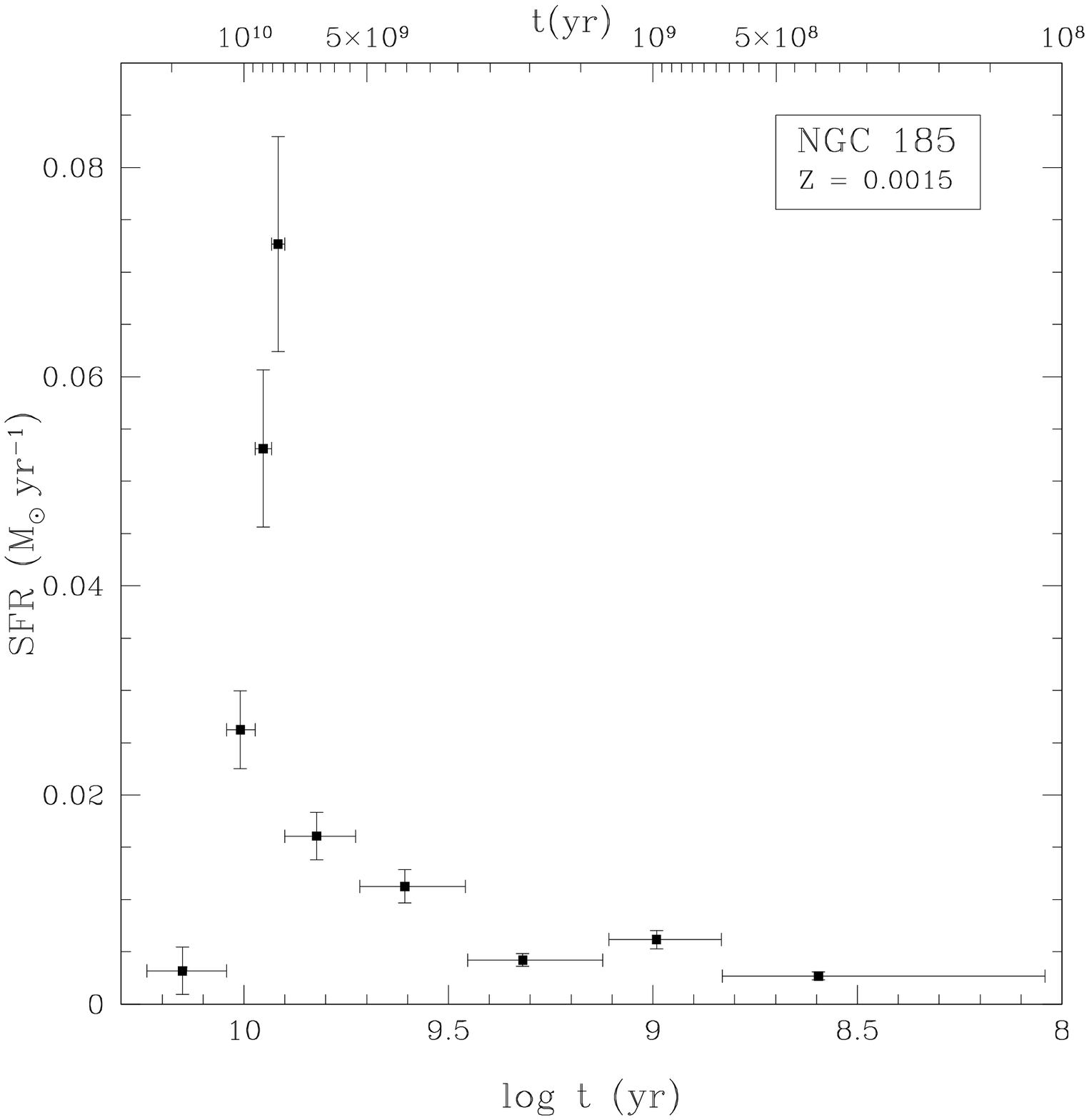,width=88mm}
}}
\caption{SFHs in the central $6\rlap{.}^\prime5\times6\rlap{.}^\prime5$ regions
of NGC\,147 ({\it left}) and NGC\,185 ({\it right}), from LPV counts assuming
a constant metallicity.}
\label{fig:fig5}
\end{figure*}

The SFR as a function of look-back time (age) in NGC\,147 and NGC\,185 is
shown in figure \ref{fig:fig5}. The horizontal bars represent the age bins.

% FIGURE 5

A peak of star formation occurred in NGC\,147 at 7 Gyr ago ($\log t=9.85$; left panel in figure \ref{fig:fig5}), when the SFR reached a level of $0.022\pm0.004$ M$_\odot$ yr$^{-1}$ on average over an interval of $\sim1.3$ Gyr. The SFR then dropped to $\sim0.01$ M$_\odot$ yr$^{-1}$ between 4--6 Gyr
ago. After a slight increase in SFR to $\sim0.014$ M$_\odot$ yr$^{-1}$ it continues with 0.003--0.005 M$_\odot$ yr$^{-1}$. Of the total stellar mass
in NGC\,147 ($1.16\times10^8$ M$_\odot$), 51\% was formed during the period of
intense star formation from 11 to 6.5 Gyr ago.

In NGC\,185 (right panel in figure \ref{fig:fig5}), the SFR shows an abrupt
and brief enhancement 8.3 Gyr ago, reaching $0.062\pm0.009$ M$_\odot$
yr$^{-1}$. After this, the SFR dropped precipitously; by 200 Myr ago it
continued at a rate of 0.006--0.008 M$_\odot$ yr$^{-1}$. Of the total stellar mass in NGC\,185 ($2.42\times10^8$ M$_\odot$),
54\% was formed between 10.2 and 7.6 Gyr ago. This is a similar fraction to
that in NGC\,147 but formed in roughly half the time.

%------------------------------------------------------------------------- 4.2
\subsection{Assuming different metallicities for the old and young stellar
populations}
\label{subsec:subsec4.2}

As a galaxy ages, the metallicity of the ISM -- and hence that of new
generations of stars -- changes as a result of nucleosynthesis and feedback
from dying stars. So we expect older stars to have formed in more metal poor
environments than younger stars have. To examine the effect of chemical
evolution on the SFH we derive, we also considered the case in which the
metallicity is lower than that assumed in section \ref{subsec:subsec4.1} for
older stars and higher for younger stars.

\begin{figure*}
\centerline{\hbox{
\epsfig{figure=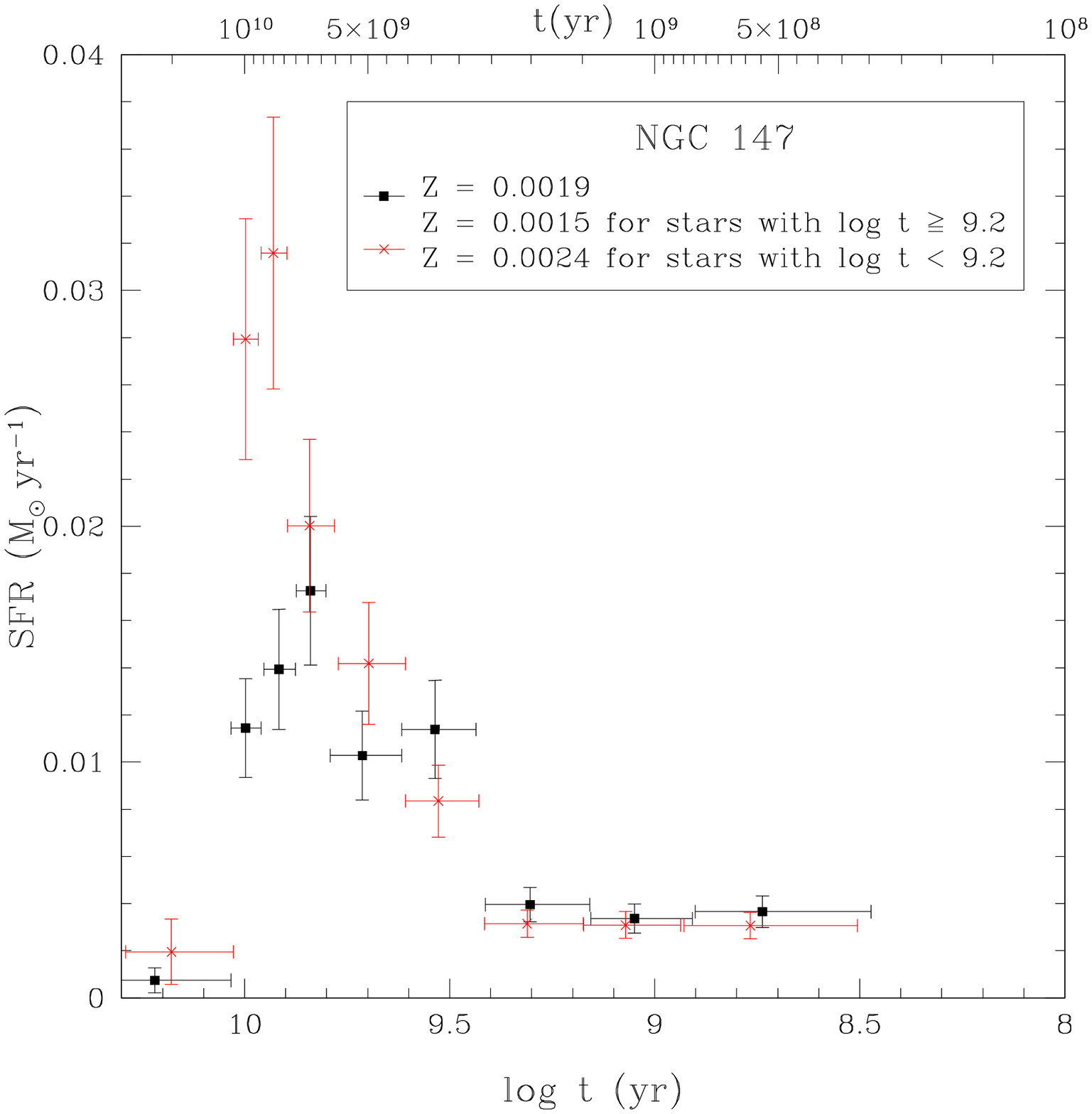,width=88mm}
\epsfig{figure=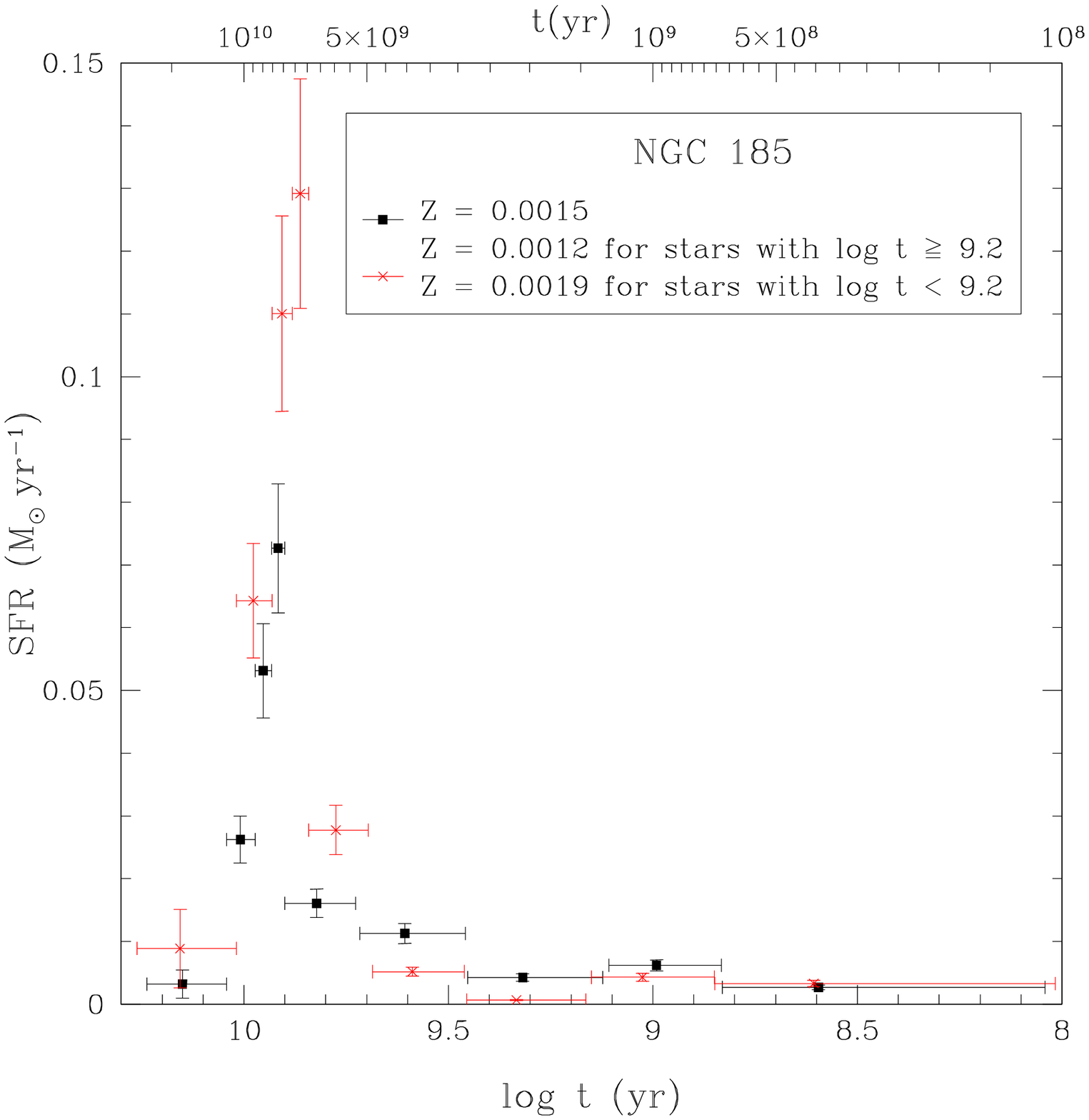,width=88mm}
}}
\caption{The effect on the SFH of a time-varying metallicity. ({\it Left:})
NGC\,147 with a constant $Z=0.0019$ (black squares), or $Z=0.0015$ for
$\log t\geq 9.2$ and $Z=0.0024$ for $\log t<9.2$ (red crosses); ({\it Right:})
NGC\,185 with a constant $Z=0.0015$ (black squares), or $Z=0.0012$ for
$\log t\geq 9.2$ and $Z=0.0019$ for $\log t<9.2$ (red crosses).}
\label{fig:fig6}
\end{figure*}

Reported metallicities span the broad range of $0.0012<Z<0.0076$ for NGC\,147
and $0.0008<Z<0.0030$ for NGC\,185. We here adopt $Z=0.0015$ for
$\log t\geq 9.2$ and $Z=0.0024$ for $\log t<9.2$, in NGC\,147 (as opposed to a
constant $Z=0.0019$), and $Z=0.0012$ and $Z=0.0019$ for these age groups in
NGC\,185 (cf.\ $Z=0.0015$). Figure \ref{fig:fig6} shows the resulting SFHs.

% FIGURE 6

Two effects are obvious. Firstly, the star formation peak becomes almost twice
as high, while more recent star formation is diminished. Secondly, the peak of
star formation shifts in age. Curiously, it shifts towards older ages in
NGC\,147 but towards more recent times in NGC\,185. The total stellar masses
obtained from integration of the SFH also become larger: $1.81\times10^8$
M$_\odot$ in NGC\,147 and $4.70\times10^8$ M$_\odot$ in NGC\,185.

%------------------------------------------------------------------------- 4.3
\subsection{Overall effect of metallicity}
\label{subsec:subsec4.3}

We here expand on the analysis of the effects on the SFH resulting from the
assumptions regarding the metallicity. The graphs and tables with the model
relations can be found in the appendix. In figure \ref{fig:fig7} we plot the
results from a wider range of assumed (constant) metallicity. Depending on the
time in the history of the galaxy at which such metallicity may have been the
norm, one could selectively read off the SFR accordingly.

% FIGURE 7
\begin{figure*}
\centerline{\hbox{
\epsfig{figure=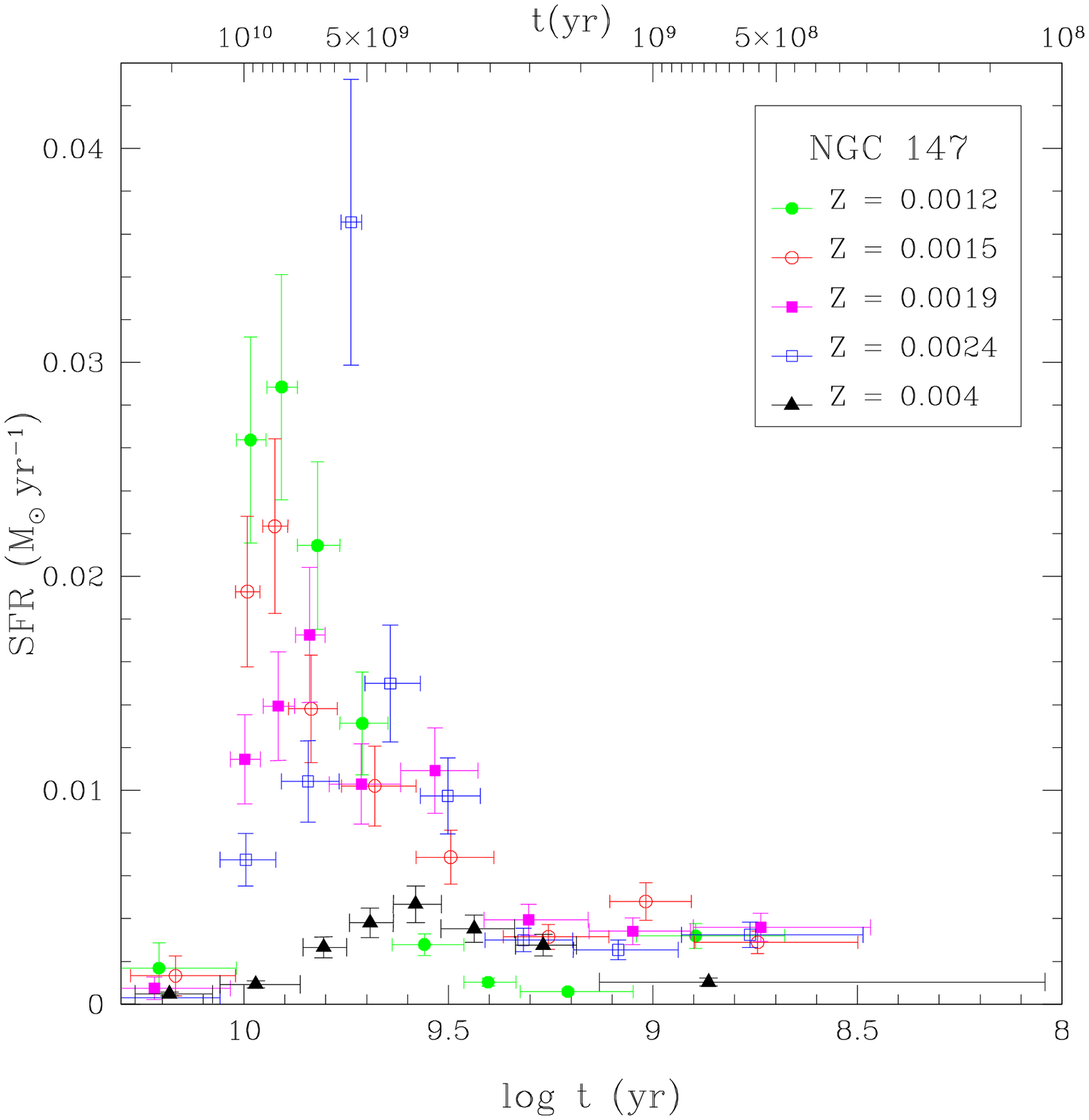,width=88mm}
\epsfig{figure=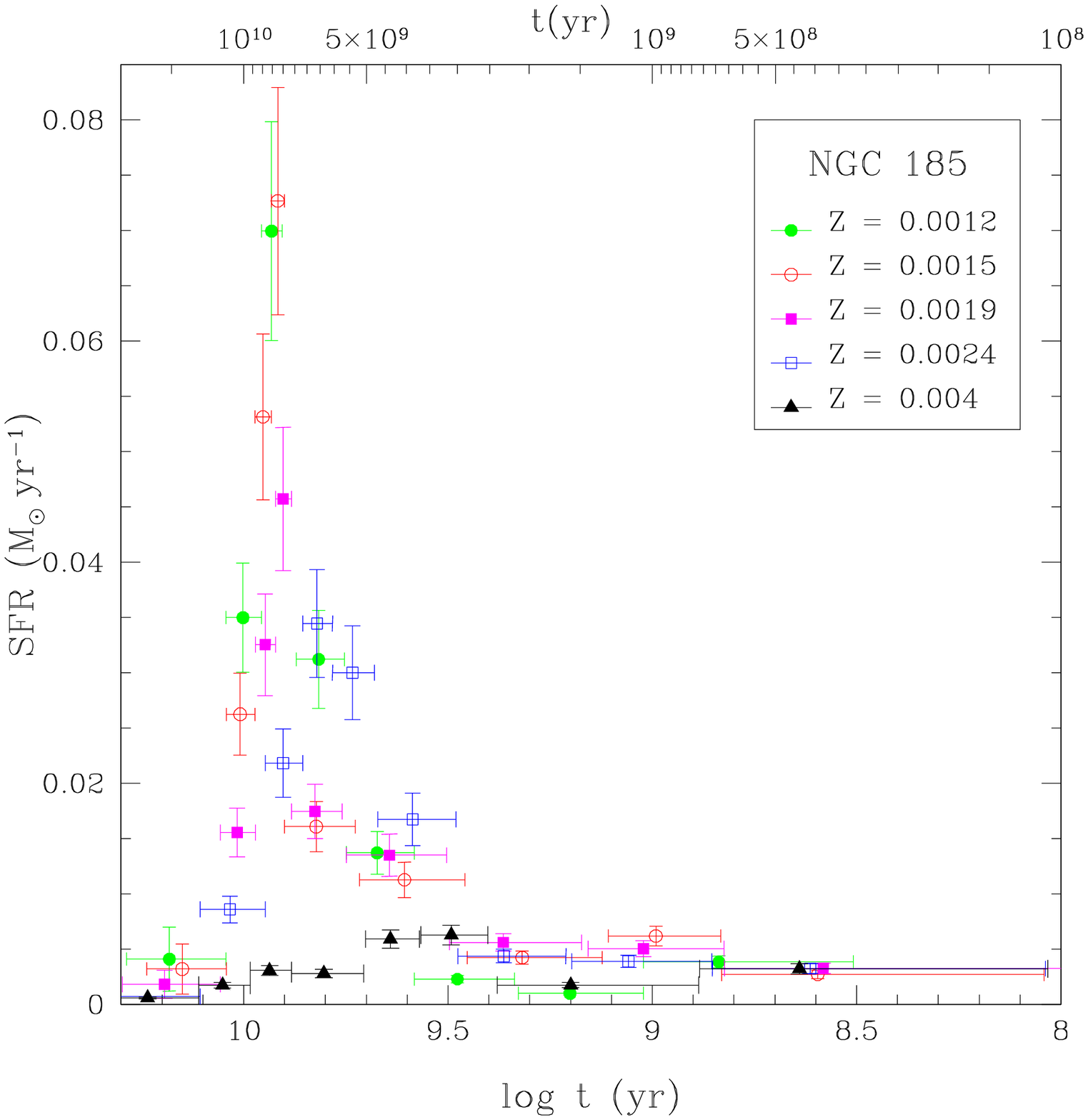,width=88mm}
}}
\caption[]{SFHs for different (constant) metallicity, for NGC\,147
({\it left}) and NGC\,185 ({\it right}).}
\label{fig:fig7}
\end{figure*}

We confirm that the peak of star formation would shift towards older ages for
lower metallicity. The effect is especially strong at the highest metallicity
we consider here, $Z=0.004$ (which is typical of the ISM within the Small
Magellanic Cloud). Such high metallicities are not expected to be reached
within either NGC\,147 or NGC\,185, except perhaps for the youngest stars in
NGC\,147. We do not expect to see the dramatic drop in SFR at old ages as a
result of such high metallicity. So our results do not change qualitatively
when changing the assumed metallicity.

% TABLE 4
\begin{table}
\caption[]{Total stellar mass for different (constant) metallicity.}
\begin{tabular}{lcc}
\hline\hline
$Z$      & $M_{\rm NGC\,147}$ (M$_\odot$) & $M_{\rm NGC\,185}$ (M$_\odot$) \\
\hline
$0.0012$ & $1.67\times10^8$             & $3.40\times10^8$             \\
$0.0015$ & $1.29\times10^8$             & $2.42\times10^8$             \\
$0.0019$ & $1.16\times10^8$             & $2.03\times10^8$             \\
$0.0024$ & $1.17\times10^8$             & $2.14\times10^8$             \\
$0.004$  & $2.97\times10^7$             & $5.05\times10^7$             \\
\hline
\end{tabular}
\label{tab:tab4}
\end{table}

% FIGURE 8
\begin{figure*}
\centerline{\hbox{
\epsfig{figure=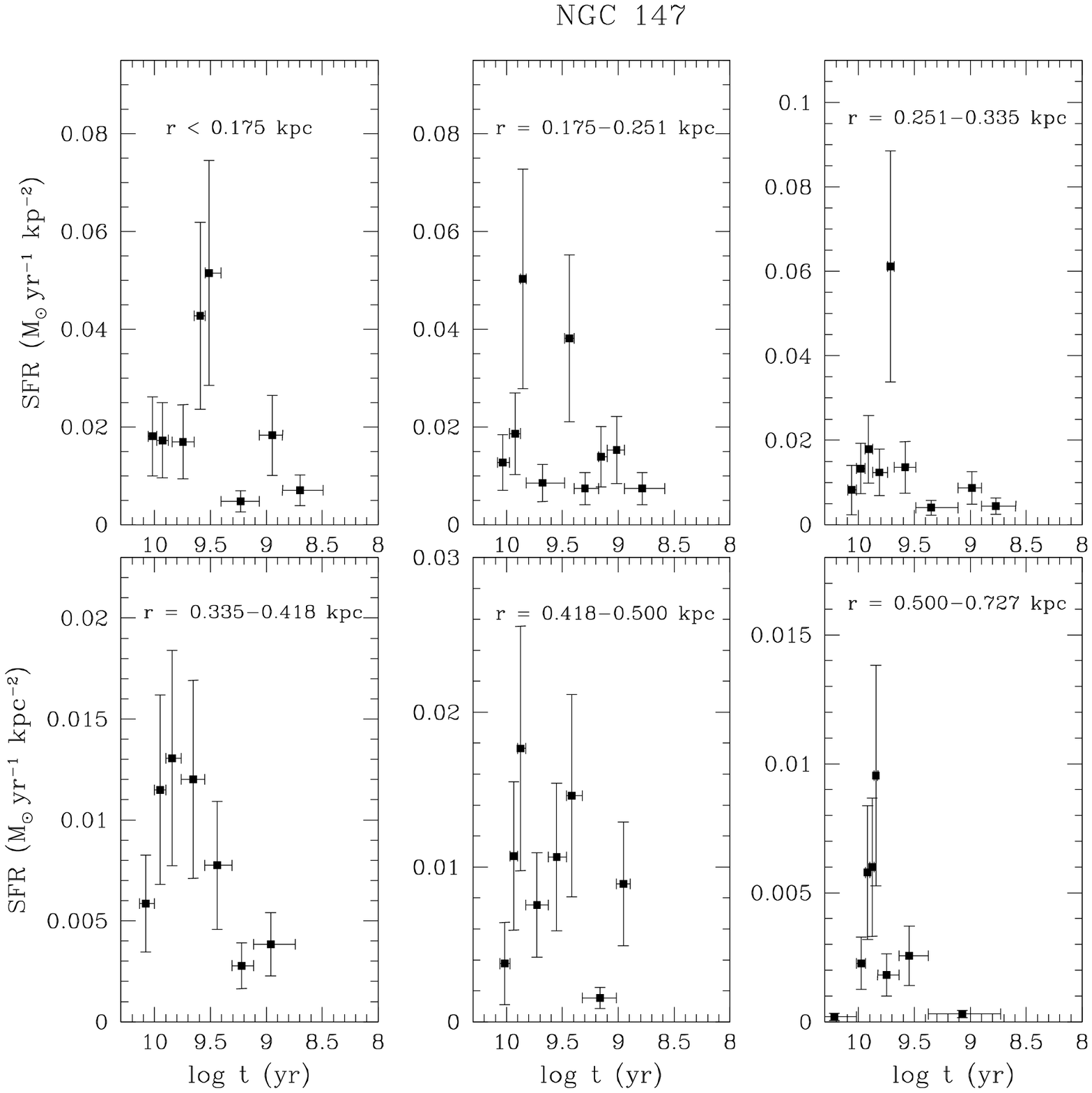,width=88mm}
\epsfig{figure=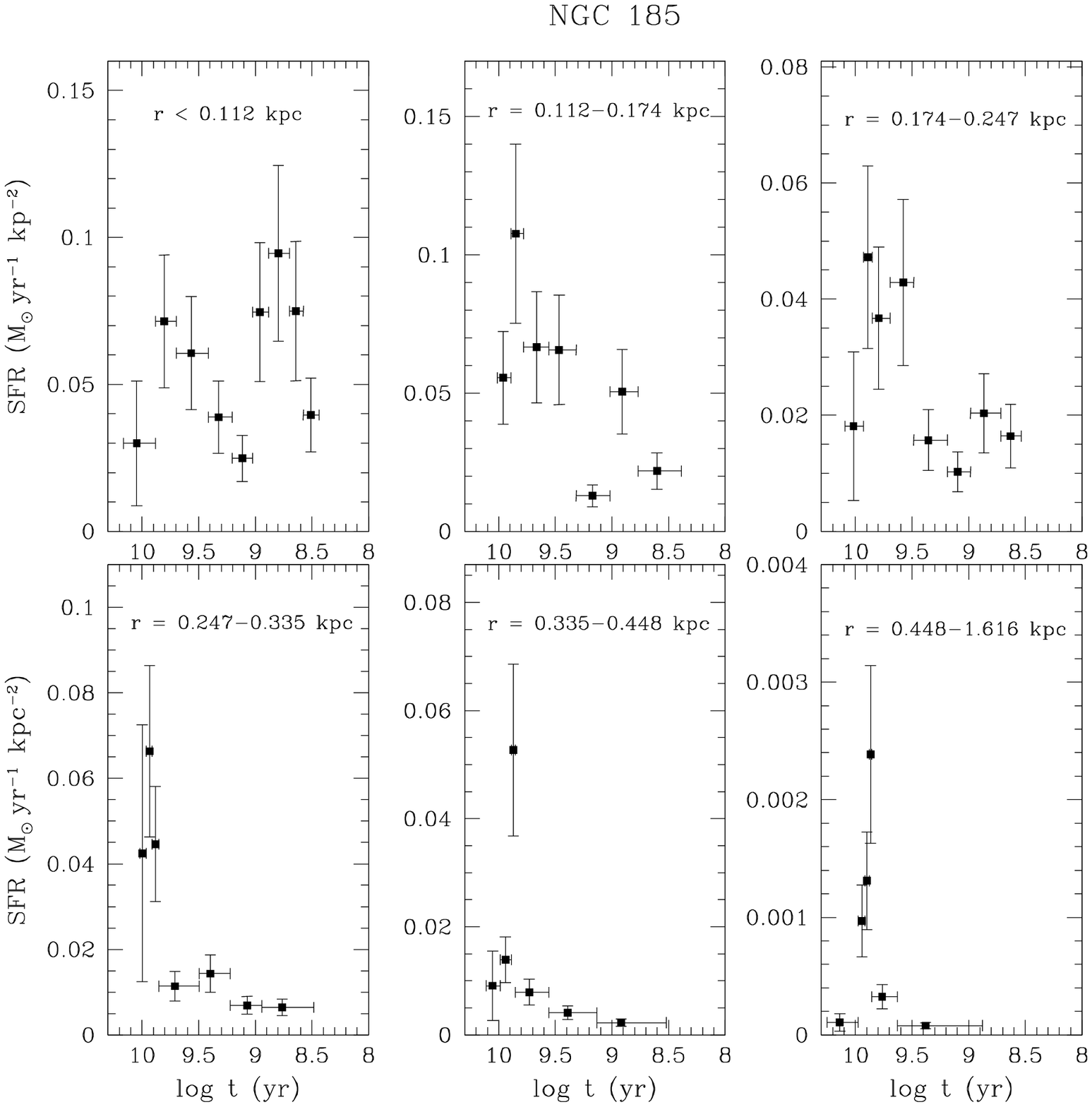,width=88mm}
}}
\caption{SFHs in NGC\,147 ({\it left}) and NGC\,185 ({\it right}) in bins at
galactocentric radii increasing from left to right and top to bottom.}
\label{fig:fig8}
\end{figure*}

The total stellar mass does depend on metallicity (see table \ref{tab:tab4});
as the metallicity increases the total stellar mass decreases. However, this
change is very small, $\lsim\,10$\%, in the range $Z=0.0015$--0.0024.

%------------------------------------------------------------------------- 4.4
\subsection{Galactocentric radial gradients of the SFHs}
\label{subsec:subsec4.4}

Besides the history of the SFR, variations across the galaxy can elucidate on
the mechanisms that governed the formation and evolution of the galaxy over
cosmological times. The suggestion of a metallicity gradient in NGC\,185 is
one such example; here we explore whether the SFH has varied with location
within the galaxy, specifically whether there is a galactocentric radial
gradient. The results are presented in figure \ref{fig:fig8}.

Roughly speaking, the peak in SFR diminishes by a factor $\sim5$ between the
innermost and outermost regions ($r\sim0.6$ kpc). Note that the tidal tails of
NGC\,147 extend $\sim2^\circ$ to either side, i.e.\ well beyond the region we
probe here. The most striking variation in SFH that we see is in NGC\,185,
where recent star formation ($t<$ Gyr or $\log t<9$) is very prominent in the
inner 0.1 kpc but completely lacking beyond 0.5 kpc. This may be related to
the strong central concentration we noted in figure \ref{fig:fig1}. NGC\,147
exhibits a more uniform SFH, possibly related to the more diffuse distribution
of AGB stars in figure \ref{fig:fig1}.

As we mentioned at the end of section 3, the dominant error on SFR is not photometric; the photometric errors are very small as the measurements are well above the
completeness limit of the surveys, but these are average values of variable
stars and the typical uncertainty can be as much as 0.1 mag. However, as can
be seen in figures 2 and 3 the spread in magnitude is very much larger (a few
magnitude) and the error on the individual stars is negligible compared to the
bins in age. This is confirmed by looking at the resulting SFHs, in figures
5-7, some of which show variation between adjacent bins that is larger than
the errorbars on those bins; this could not have happened if the photometric
uncertainties were large, especially as it is seen at some of the older bins
which are associated with the fainter stars that would be expected to have
larger photometric errors.

%=========================================================================== 5
\section{Discussion: a different evolution of NGC\,147 compared to NGC\,185}
\label{sec:sec5}

In NGC\,147, the small population of early-AGB stars led Han et al.\ (1997) to
infer the presence of intermediate aged (several Gyr) stars. They did not see
any main sequence star in NGC\,147, from which they concluded that star
formation ceased about a Gyr ago. The distribution of AGB stars and Horizontal
Branch stars in NGC\,147 suggested that the younger stars are more centrally
concentrated. Han et al.\ (1997) also detected a metallicity gradient, with
the most metal-rich stars residing in the central regions. Our independent
determination of the SFH of NGC\,147 is in broad agreement with the above
findings. In particular, we find that star formation ceased $\sim300$ Myr ago.
Star formation was significant at intermediate ages, between $\sim7$--3 Gyr
ago. If we assume a higher metallicity for young stars and a lower metallicity
for older stars, then the start of the enhanced star formation shifts to older
ages, $\sim8$ Gyr ago, but star formation still had all but ceased by 300 Myr
ago (see Fig.\ \ref{fig:fig7}; $\log t=8.5$). These results are further
corroborated by the near-IR study of the inner $2^\prime\times2^\prime$ by
Davidge (2005); he found that the most luminous AGB stars in NGC\,147 are well
mixed with fainter stars, and that the most recent significant star formation
occurred $\sim3$ Gyr ago.

% FIGURE 9
\begin{figure}
\centerline{\hbox{
\epsfig{figure=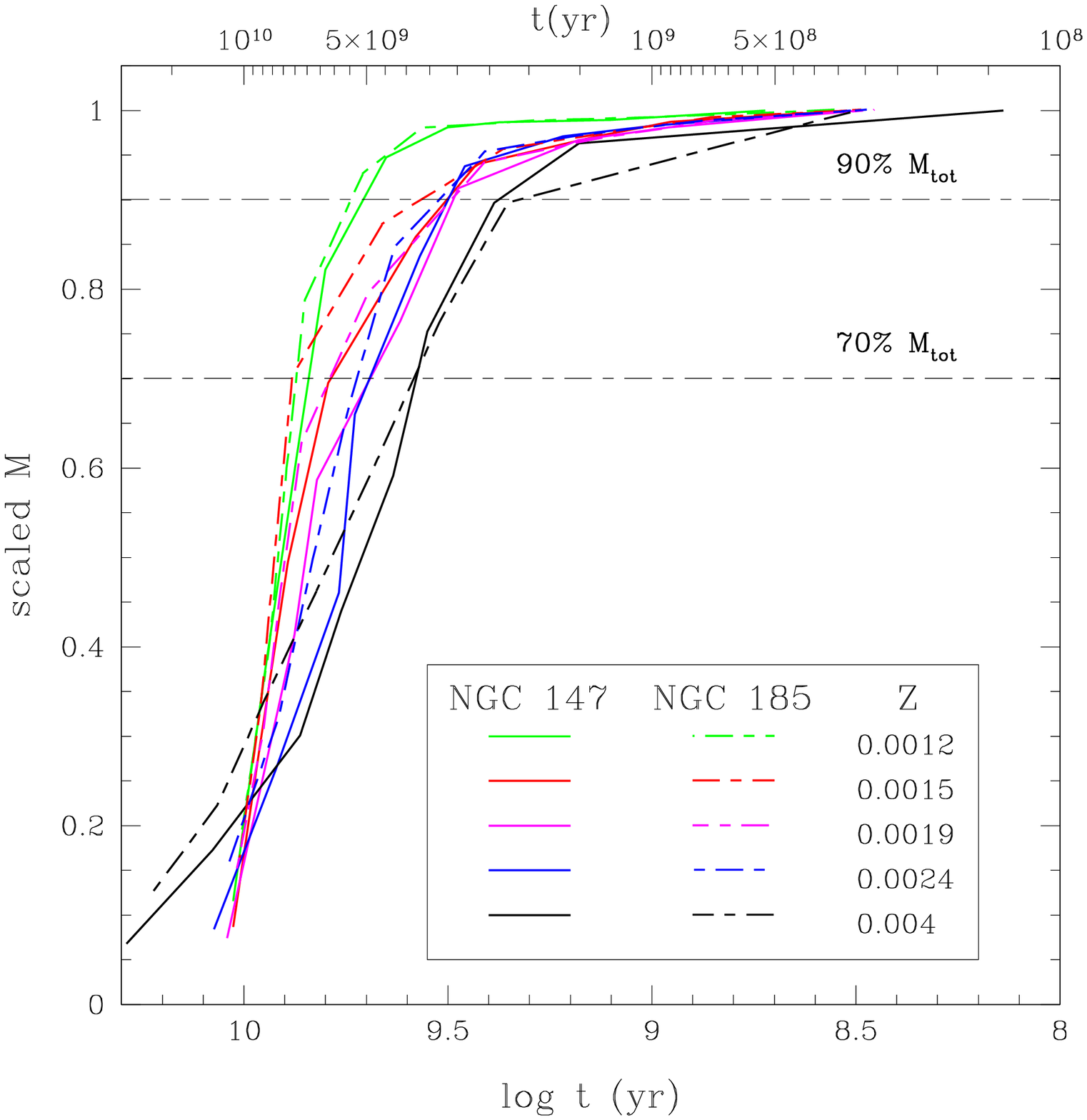,width=88mm}
}}
\caption{Cumulative SFH in the central
$6\rlap{.}^\prime5\times6\rlap{.}^\prime5$ regions of NGC =\,147 
and NGC\,185.}
\label{fig:fig9}
\end{figure}

In NGC\,185, Mart\'{\i}nez-Delgado \& Aparicio (1998) determined a metallicity
gradient within a $7\rlap{.}^\prime2\times7\rlap{.}^\prime2$ region, on the
basis of optical photometry. Additional photometry was used to derive the SFH
(Mart\'{\i}nez-Delgado, Aparicio \& Gallart 1999). They found that the bulk of
stellar mass was put in place early on, but star formation continued until at
least $\sim100$ Myr ago. They measured a mean SFR $\xi\approx8.2\times10^{-3}$
M$_\odot$ yr$^{-1}$ between 1--15 Gyr ago, and $\xi\approx6.6\times10^{-4}$
M$_\odot$ yr$^{-1}$ over the most recent Gyr. The most recent star formation is
confined to the central $150\times90$ pc$^2$. The results we obtained in an
entirely different way are remarkably consistent with theirs; star formation
sharply peaked $\sim8$ Gyr ago, but star formation lasted until $\sim200$ Myr ($\log t\approx 8.3$) $<170$ pc from the centre of NGC\,185. The only discrepancy is that our SFRs
are about twice their values. Like for NGC\,147, Davidge (2005) used near-IR
photometry of AGB stars; he found that in NGC\,185 the most recent significant
star formation occurred $\sim1$ Gyr ago, i.e.\ much more recently than in
NGC\,147.

We derived total stellar masses of $M\approx1.16\times10^8$ M$_\odot$ for
NGC\,147 and $M\approx2.42\times10^8$ M$_\odot$ for NGC\,185. The latter is in
excellent agreement with the total stellar mass of $M\sim2\times10^8$ M$_\odot$
derived by Davidge (2005) for NGC\,185 on the basis of near-IR photometry and
assuming a mass-to-light ratio $M/L=1$; Geha et al.\ (2010) adopted somewhat
larger $M/L$ values and hence obtained $M\sim5.6\times10^8$ M$_\odot$ for
NGC\,147 and $M\sim7.2\times10^8$ M$_\odot$ for NGC\,185, i.e.\ 3--4 times
higher than our estimates. Weisz et al.\ (2014a) obtained $M\sim0.6\times10^8$
M$_\odot$ for NGC\,147 and $M\sim0.7\times10^8$ M$_\odot$ for NGC\,185, but
these estimates do not include all of the galaxy and are therefore lower
limits. Their NGC\,147 field was roughly centred and covered $\approx0.23$ of
the half-light area, so we may expect the corrected total mass estimate to be
a few times higher, i.e.\ closer to $2\times10^8$ M$_\odot$; their NGC\,185
fields covered an area of similar size to the half-light area but off-centre
by 200--800 pc, so the corrected total mass estimate would again be higher by
a factor of a few, i.e.\ $\sim2\times10^8$ M$_\odot$. On balance, therefore,
our estimates of the total stellar mass seem to agree with other, independent
estimates, or perhaps be lower by a factor of a few at most. We may have
missed some mass residing outside of the area covered by the monitoring data,
or in the very oldest stars -- which do not pulsate that strongly and which
are therefore difficult to identify and account for -- but not dramatically. Moreover, since the variable star survey is incomplete due to crowding in the central field of NGC 185, this may lead to star formation rates that are underestimated.

Figure \ref{fig:fig9} shows the cumulative SFH, i.e.\ the build-up of stellar
mass over cosmological time. It is clear that the peak of star formation
occurred earlier in NGC\,185 than it did in NGC\,147. It is also clear that
the recent star formation in NGC\,185 is insignificant in terms of stellar
mass. Weisz et al.\ (2014b) argued that galaxies in which at least 70\% of the
stellar mass was in place by the time of cosmic reionization, could be
considered `fossil' galaxies. According to that definition, NGC\,147 and
NGC\,185 are not fossil galaxies, but have been actively evolving since cosmic
reionization. Weisz et al.\ (2014b) determined that 70\% of the stellar mass
was in place by $\log t=9.76$ for NGC\,147 and by $\log t=9.92$ for NGC\,185.
We find very similar timings: $\log t=9.71$ for NGC\,147 and $\log t=9.90$ for
NGC\,185. Weisz et al.\ (2015) then went on to quantify quenching of galaxy
evolution, by the time by which 90\% of the stellar mass is in place.
According to this definition, the quenching of star formation occurred around
$\log t=9.43$ in NGC\,147 and $\log t=9.56$ in NGC\,185; we find $\log t=9.55$
and $\log t=9.56$, respectively, i.e.\ sooner but similar in a relative sense.
However, this neglects the recent star formation in NGC\,185 as it accounts
for $\ll 10$\% of the stellar mass. By their definition, even the Milky Way
would be close to being quenched now. What does it mean if star formation is
quenched? What it does {\it not} mean, apparently, is that star formation is
excluded from happening.

Mart\'{\i}nez-Delgado et al.\ (1999) argued that the ISM in dEs could be
replenished by mass loss from dying earlier generations of stars, and retained
and fuel further star formation, possibly in an episodic fashion. This would
explain the recent star formation and presence of ISM in NGC\,185. The lack of
ISM in NGC\,147, which is consistent with the absence of stars younger than
300 Myr, could be due to external processes such as ram pressure due to the
gaseous halo of M\,31. Indeed, NGC\,147 is tidally distorted and must
therefore have spent some time recently in close proximity to M\,31. Geha et
al.\ (2015) suggest that the orbit of NGC\,185 has a larger pericenter as
compared to NGC\,147, allowing it to preserve radial gradients and maintain a
small central reservoir of recycled gas. They interpret the differences in
early-SFH to imply an earlier infall time into the M\,31 environment for
NGC\,185 as compared to NGC\,147. Arias et al.\ (2016) determined the likely
orbits of NGC\,147 and NGC\,185 and used them to perform $N$-body simulations
to follow their morphological evolution. They reproduced the tidal features
seen in NGC\,147 (but not in NGC\,185); their models suggest NGC\,147 and
NGC\,185 (and the Cass\,II satellite) have been a bound group for at least a
Gyr, and that their masses are $6.1\times10^8$ M$_\odot$ for NGC\,147 and
$6.6\times10^8$ M$_\odot$ for NGC\,185, similar to the aforementioned values.

It may also be possible, however, that NGC\,185 became more centrally
concentrated early on, preventing the removal of the ISM from its centre,
whereas a more diffuse NGC\,147 would have been more susceptible to tidal and
ram pressure stripping even if it were on a similar orbit.

%=========================================================================== 6
\section{Summary of conclusions}
\label{sec:sec6}

We have applied the novel method of Javadi et al.\ (2011) using long-period
variable AGB stars to derive the SFH in the M\,31 satellite dwarf elliptical
galaxies NGC\,147 and NGC\,185. Our main findings are:
\begin{itemize}
\item[$\bullet$]{Star formation started earlier in NGC\,185 than in NGC\,147,
peaking around 8.3 Gyr ago in NGC\,185 and 7 Gyr ago in NGC\,147.}
\item[$\bullet$]{Star formation continued in NGC\,185 over the past 6 Gyr,
albeit at a much lower rate, until as recent as 200 Myr ago in the centre, but
it ceased much earlier in the outskirts; in NGC\,147, on the other hand, while
star formation was significant between 3--6 Gyr ago, no star formation is seen
for the past 300 Myr, and stars of all ages are more uniformly distributed.}
\item[$\bullet$]{The total stellar mass is $>2.42\times10^8$ M$_\odot$ for
NGC\,185 and $>1.16\times10^8$ M$_\odot$ for NGC\,147; the true values are
unlikely to be higher by more than a factor of a few.}
\item[$\bullet$]{Of the total stellar mass, 70\% (90\%) was in place by
$\log t=9.90$ (9.56) in NGC\,185 and by $\log t=9.71$ (9.55)} in NGC\,147.
\item[$\bullet$]{Our conclusions were obtained completely independently, using
different data and a different method, yet they are corroborated by previous
work.}
\item[$\bullet$]{We thus confirm the possibility of a scenario in which
NGC\,147 was accreted into the environs of M\,31 later than NGC\,185, but
that its current orbit takes it closer to M\,31 explaining its lack of ISM,
lack of recent star formation, and its tidal distortions as opposed to the
ISM and recent star formation in the centre of NGC\,185. Alternatively,
NGC\,185 may have become more centrally concentrated early on, while NGC\,147
has not.}
\end{itemize}

%=============================================================================
\section*{Acknowledgments}
We are grateful for financial support by the Royal Society under grant No.\
IE130487. We also thank the referee for her/his useful report which
prompted us to improve the manuscript.

%=============================================================================

%=========================================================================== A
\appendix
%-------------------------------------------------------------------------- A1
\section{Supplementary material}
\label{app:app}

Here we present plots and parameterizations of the mass--luminosity, mass--age
and mass--pulsation duration relations derived from the theoretical models of
Marigo et al.\ (2008) and used in our analysis of the star formation histories
of NGC\,147 and NGC\,185. The methodology and the case of $Z=0.0019$ and
$\mu=24.15$ mag are presented in the main body of the paper (section 3, figure
4 and tables 1--3).

% FIGURE A1
\begin{figure*}
\centerline{\hbox{
\epsfig{figure=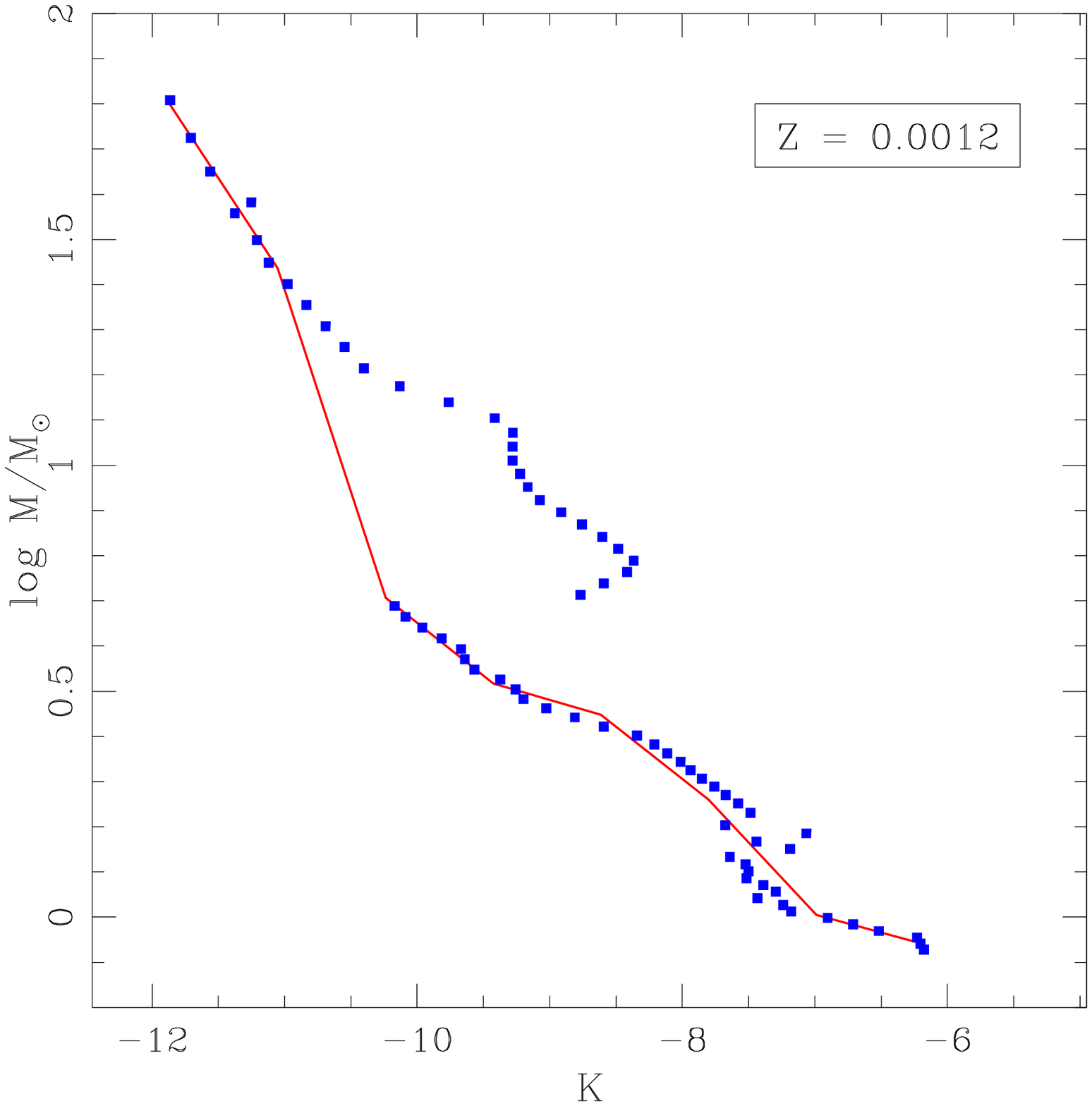,width=58mm}
\epsfig{figure=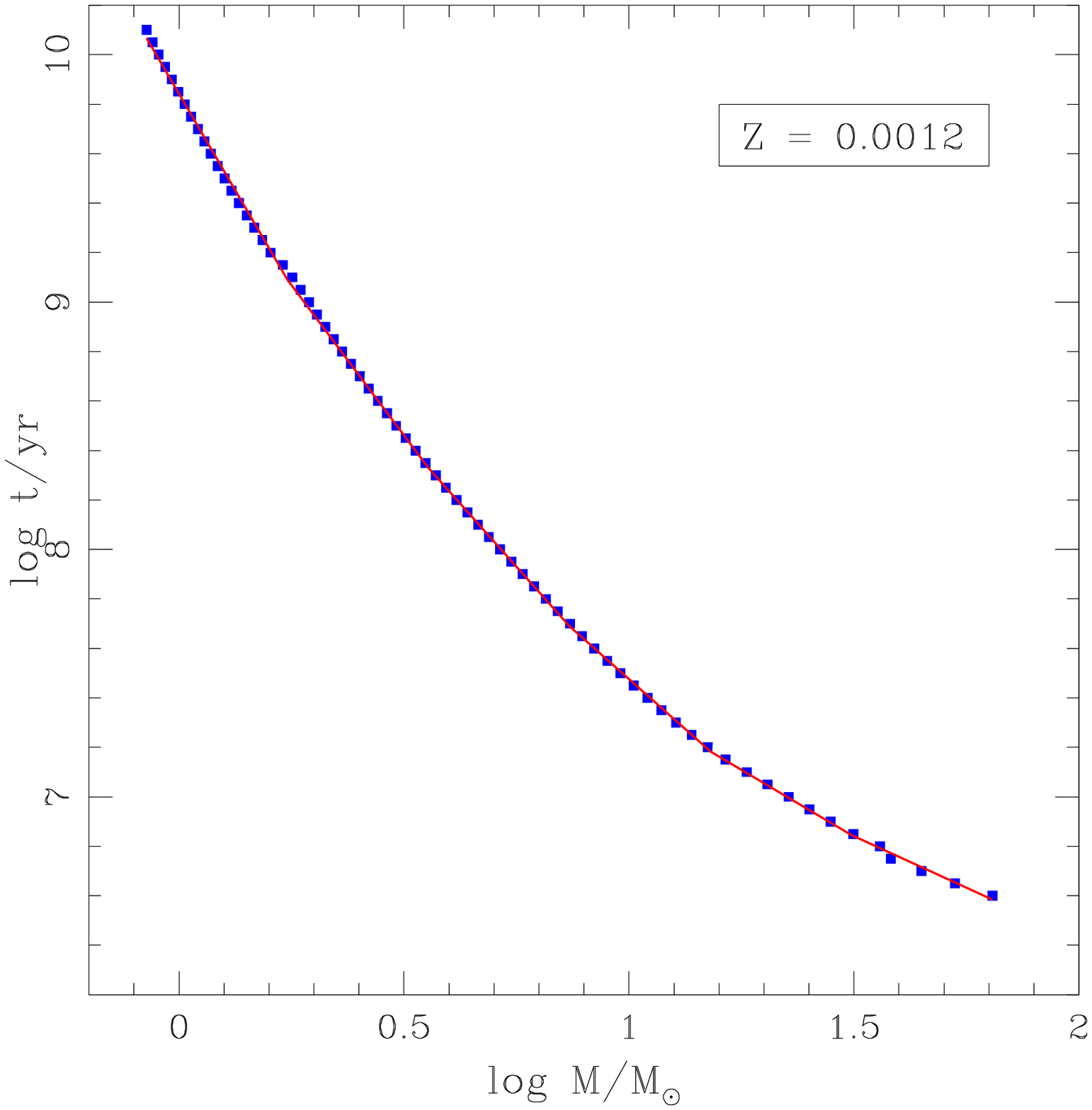,width=58mm}
\epsfig{figure=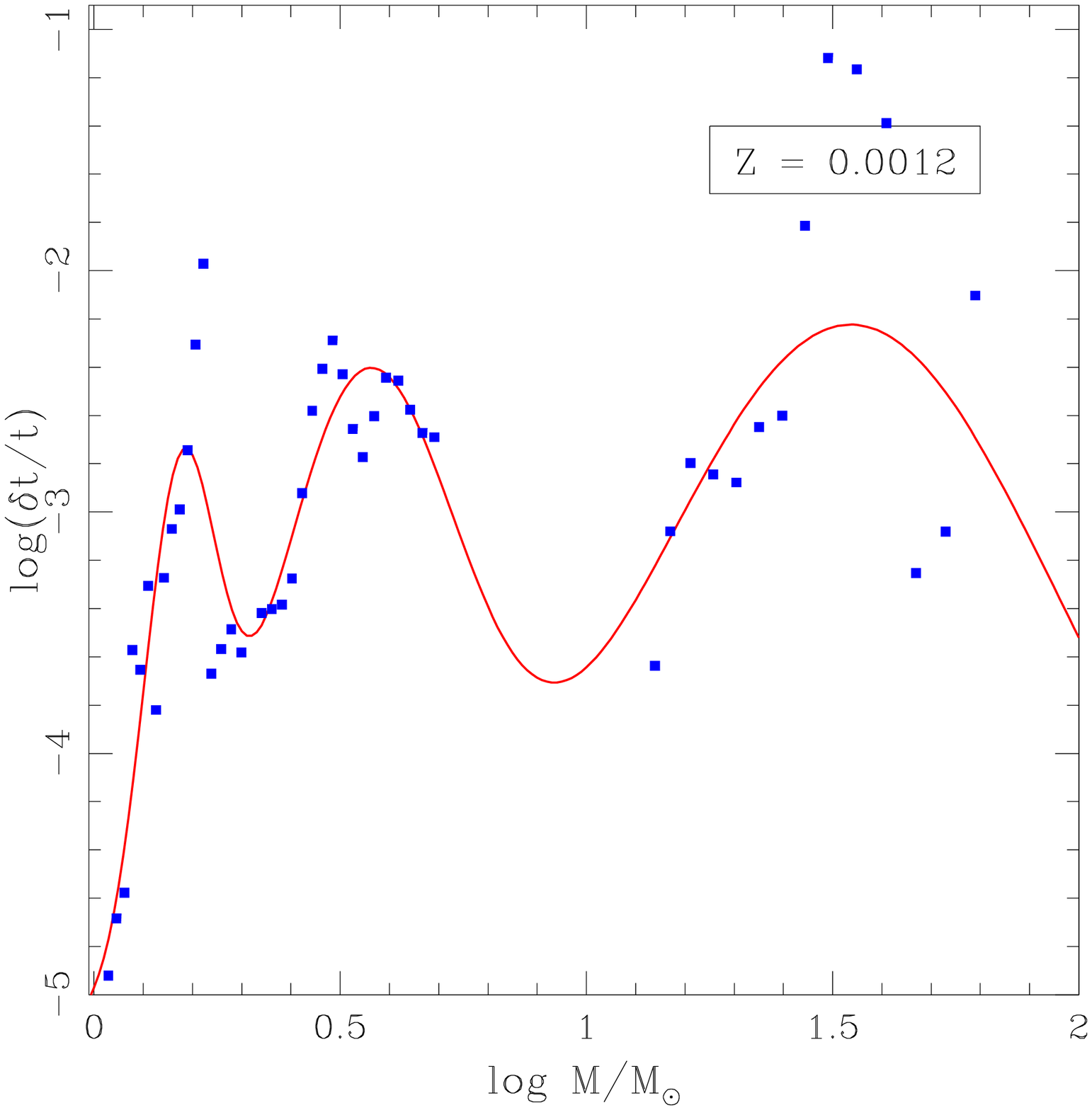,width=58mm}
}}
\caption[]{The dots refer to models from Marigo et al. (2008); ({\it Left:}) Mass--luminosity relation (in the $K$ band) for
$Z=0.0012$ and a distance modulus of $\mu=0$ mag. The solid lines are linear
spline fits, in which the function is interpolated over the super-AGB phase to
massive red supergiants. ({\it Middle:}) Same, for the mass--age relation.
({\it Right:}) Same, for the mass--pulsation duration relation. The points
show the ratio of pulsation duration to age, versus mass; the solid lines are
multiple-Gaussian fits, interpolated over the super-AGB regime.}
\end{figure*}

% FIGURE A2
\begin{figure*}
\centerline{\hbox{
\epsfig{figure=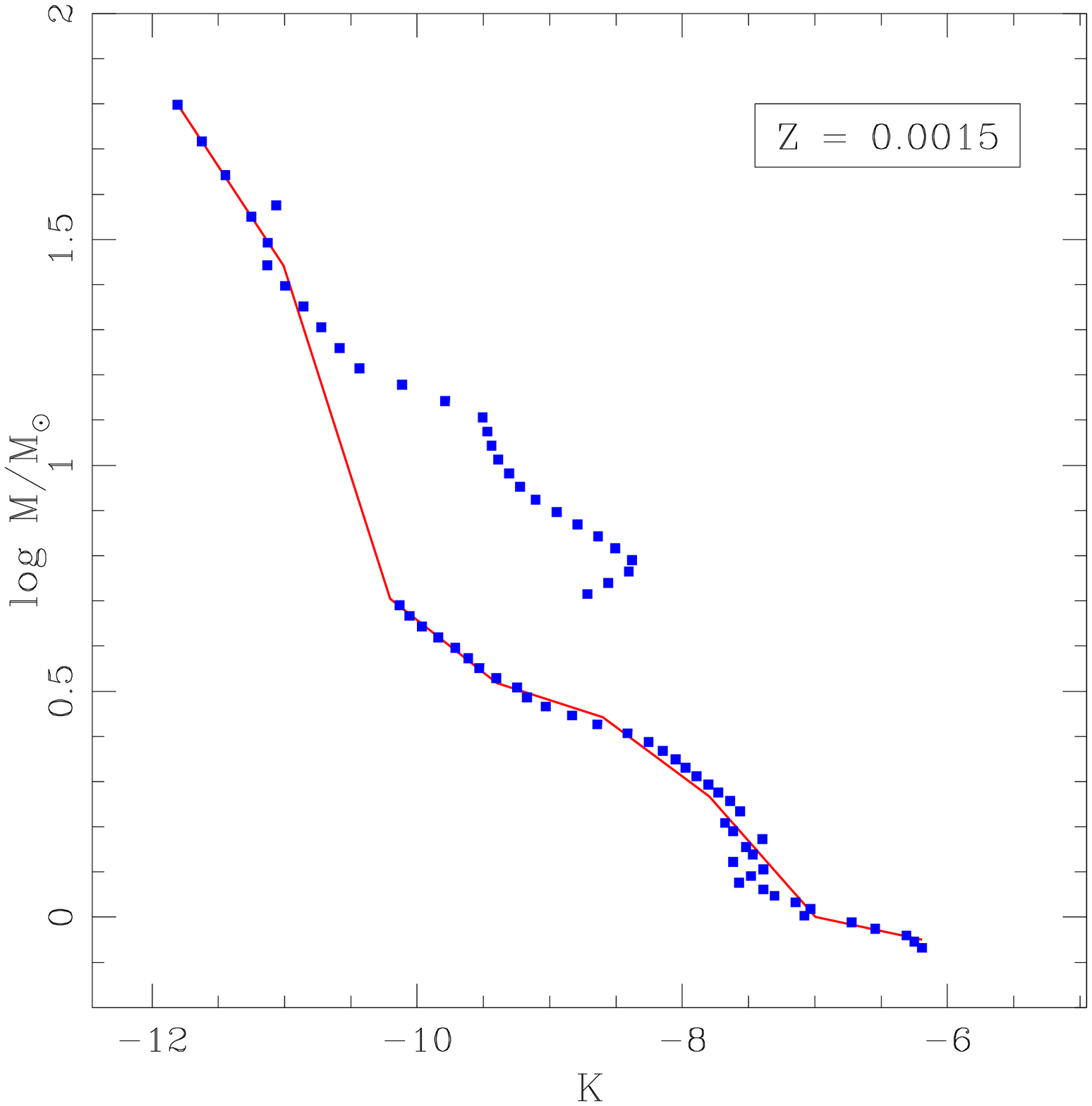,width=58mm}
\epsfig{figure=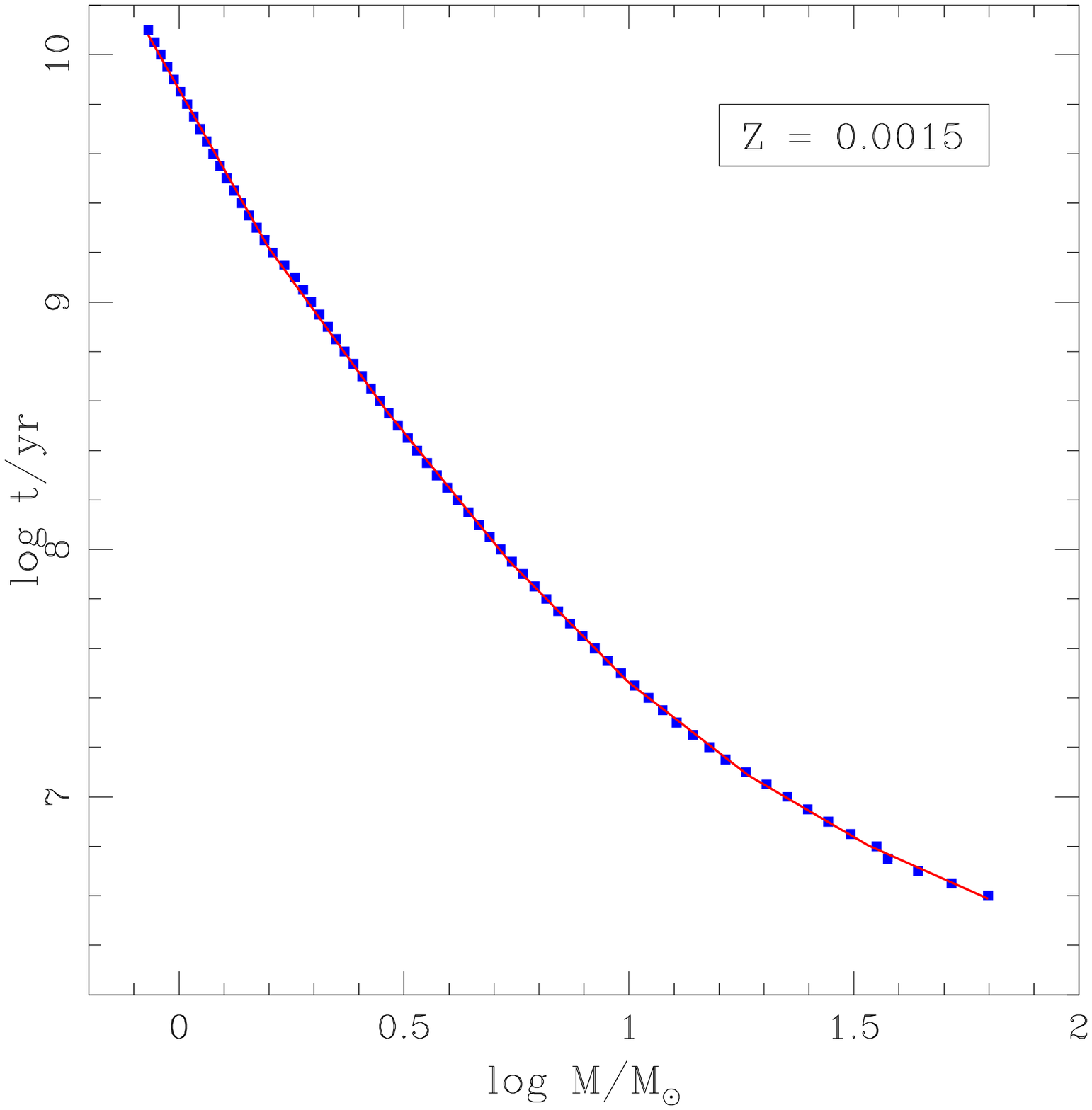,width=58mm}
\epsfig{figure=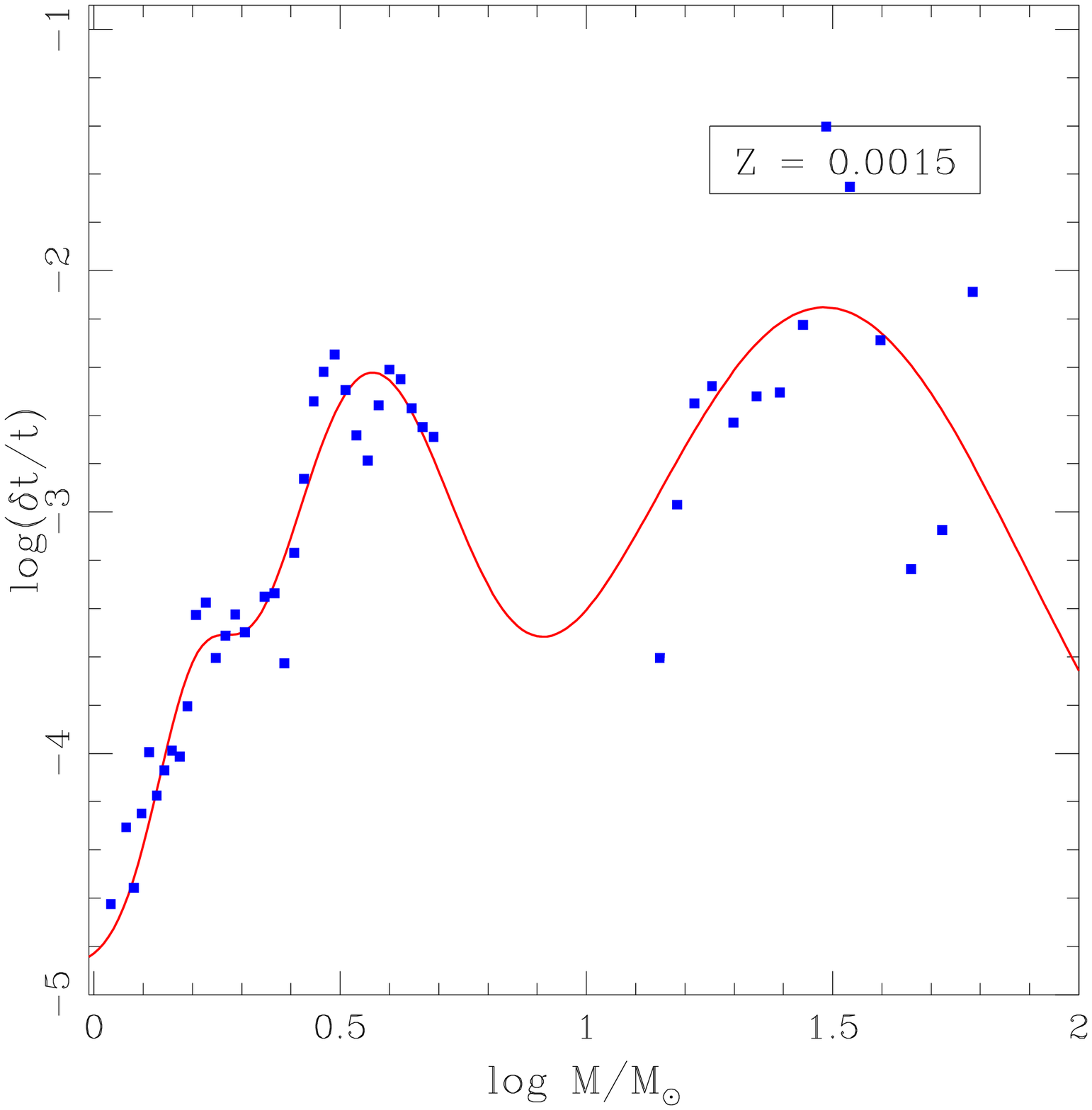,width=58mm}
}}
\caption[]{Same as Figure A1 for $Z=0.0015$.}
\end{figure*}

% FIGURE A3
\begin{figure*}
\centerline{\hbox{
\epsfig{figure=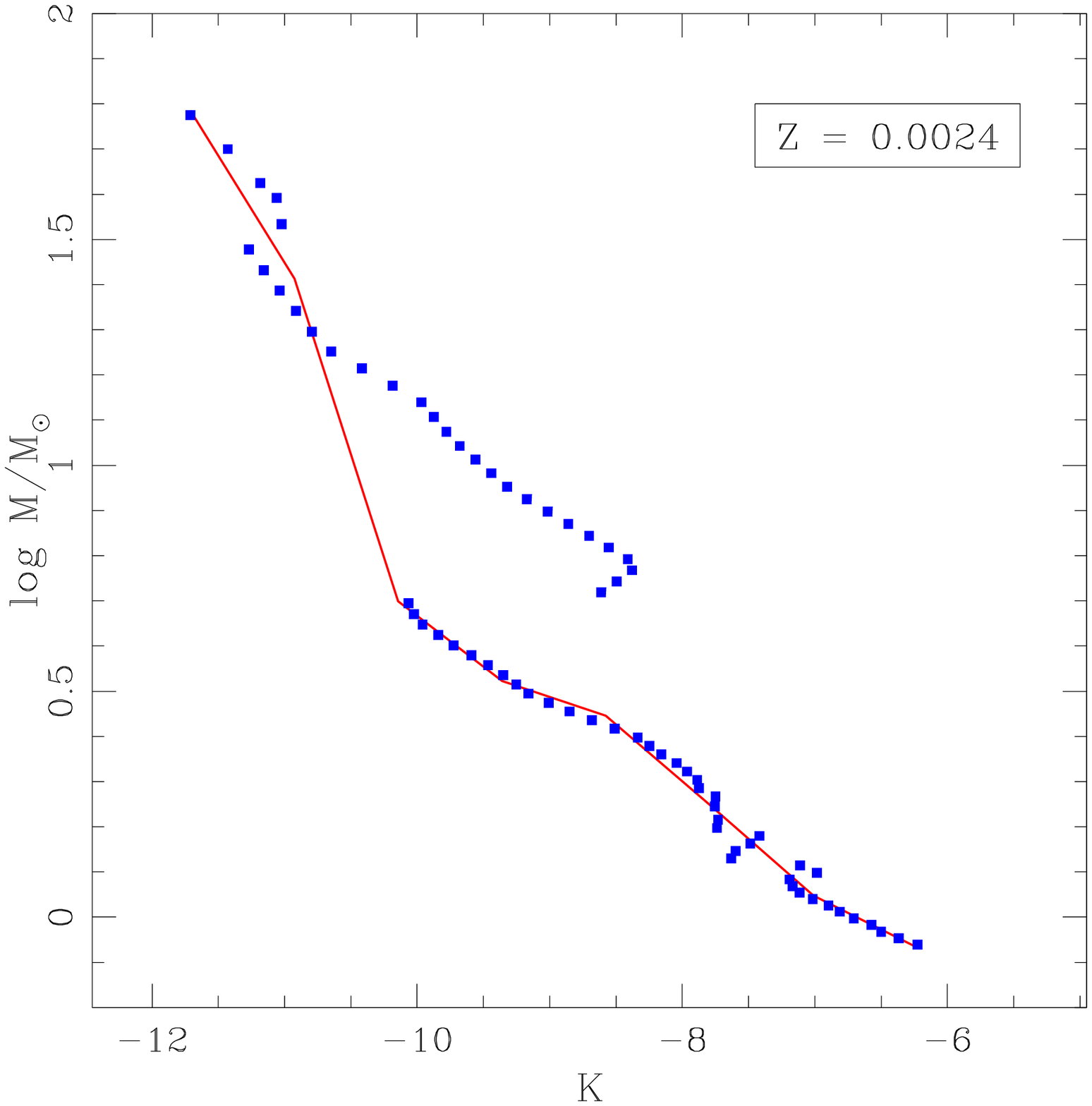,width=58mm}
\epsfig{figure=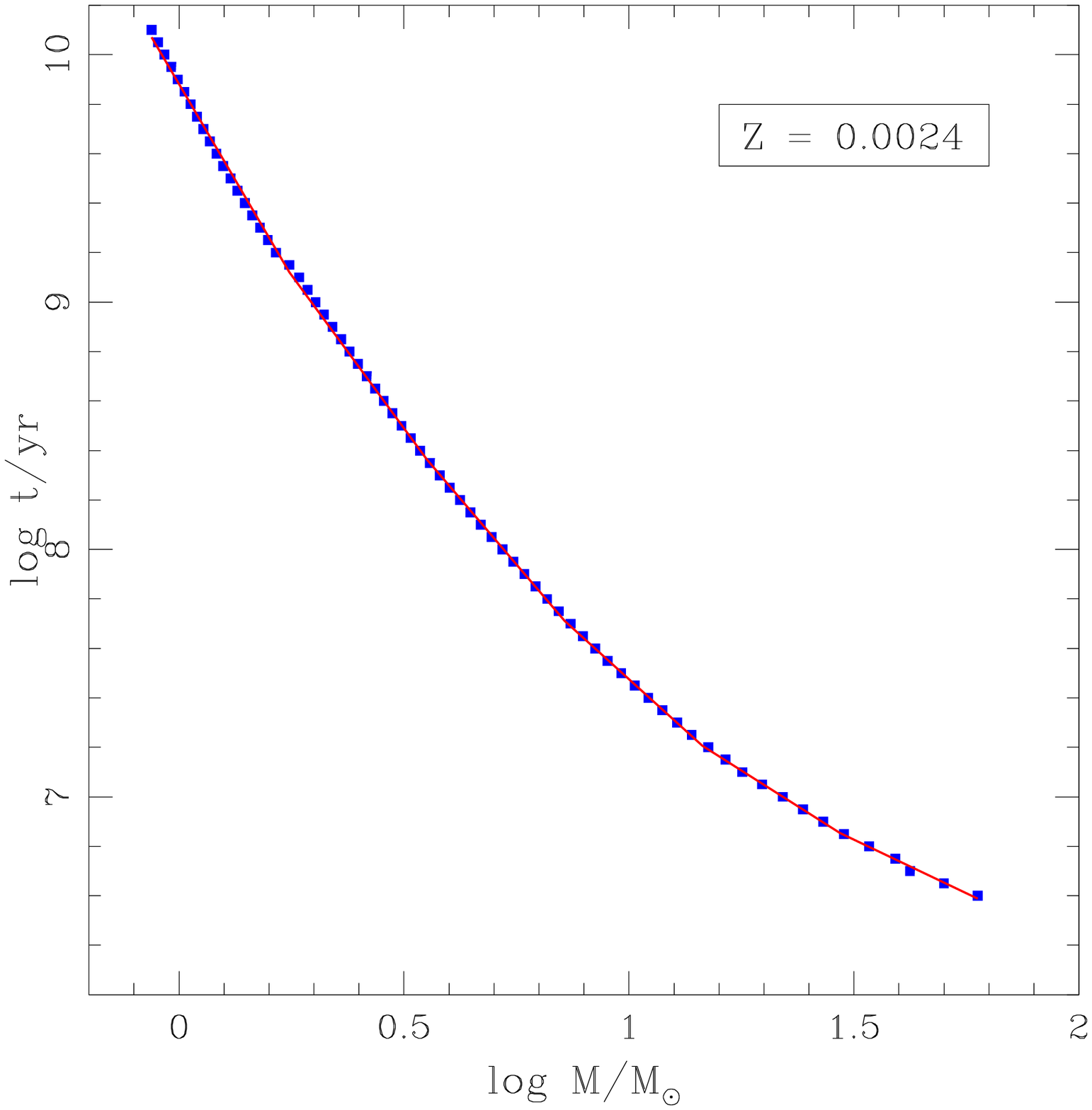,width=58mm}
\epsfig{figure=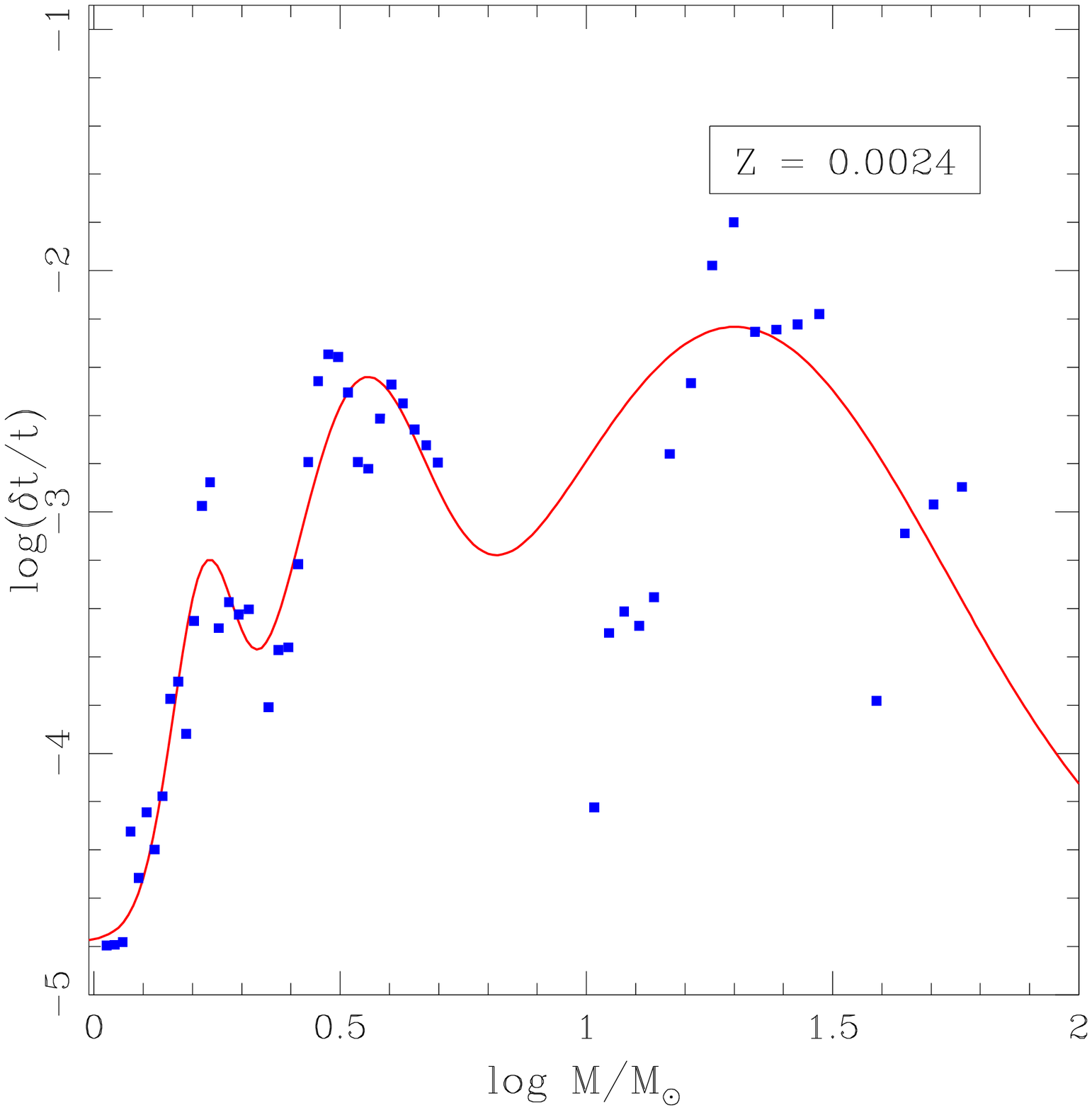,width=58mm}
}}
\caption[]{same as Figure A1 for $Z=0.0024$.}
\end{figure*}

% FIGURE A4
\begin{figure*}
\centerline{\hbox{
\epsfig{figure=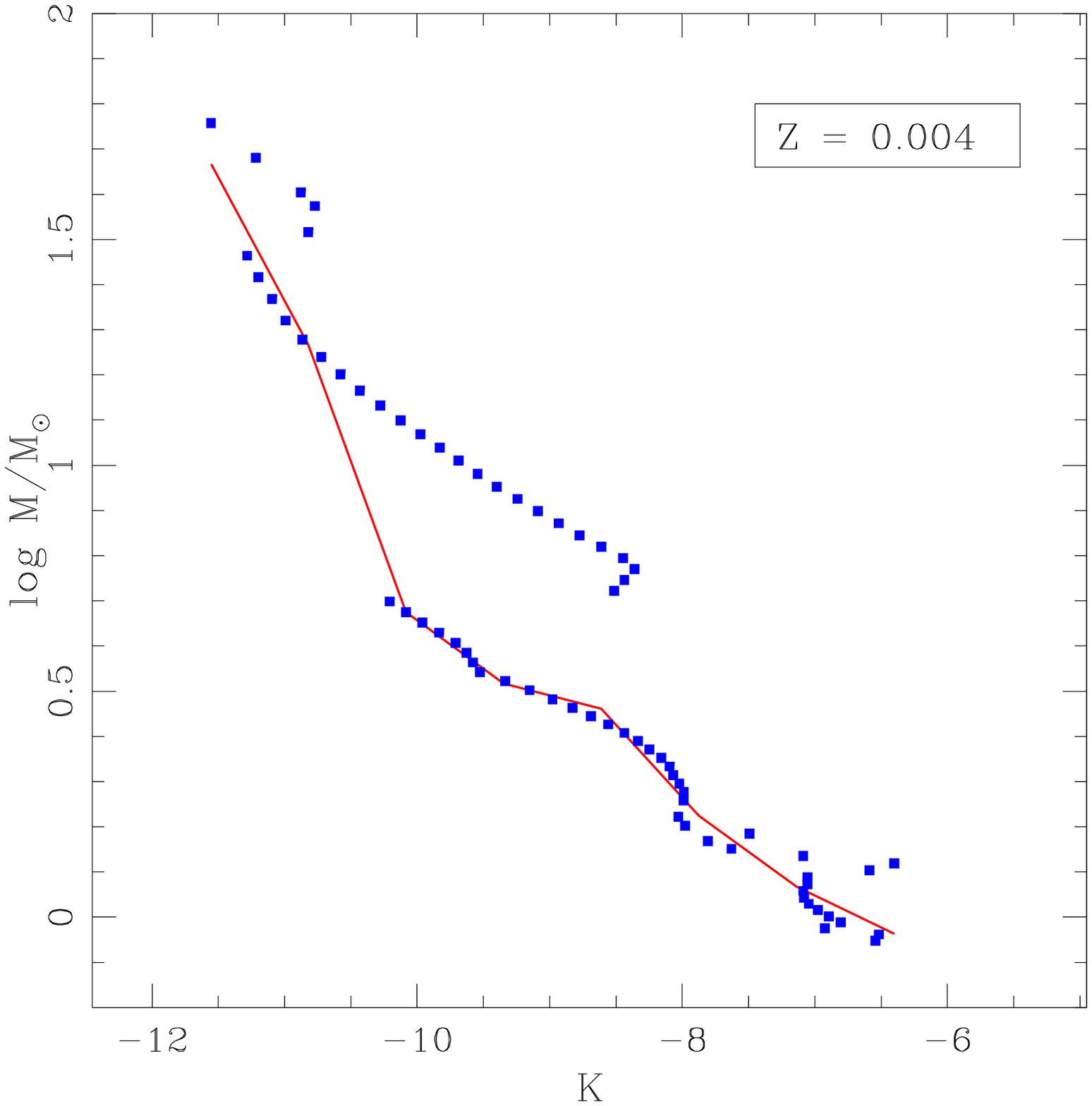,width=58mm}
\epsfig{figure=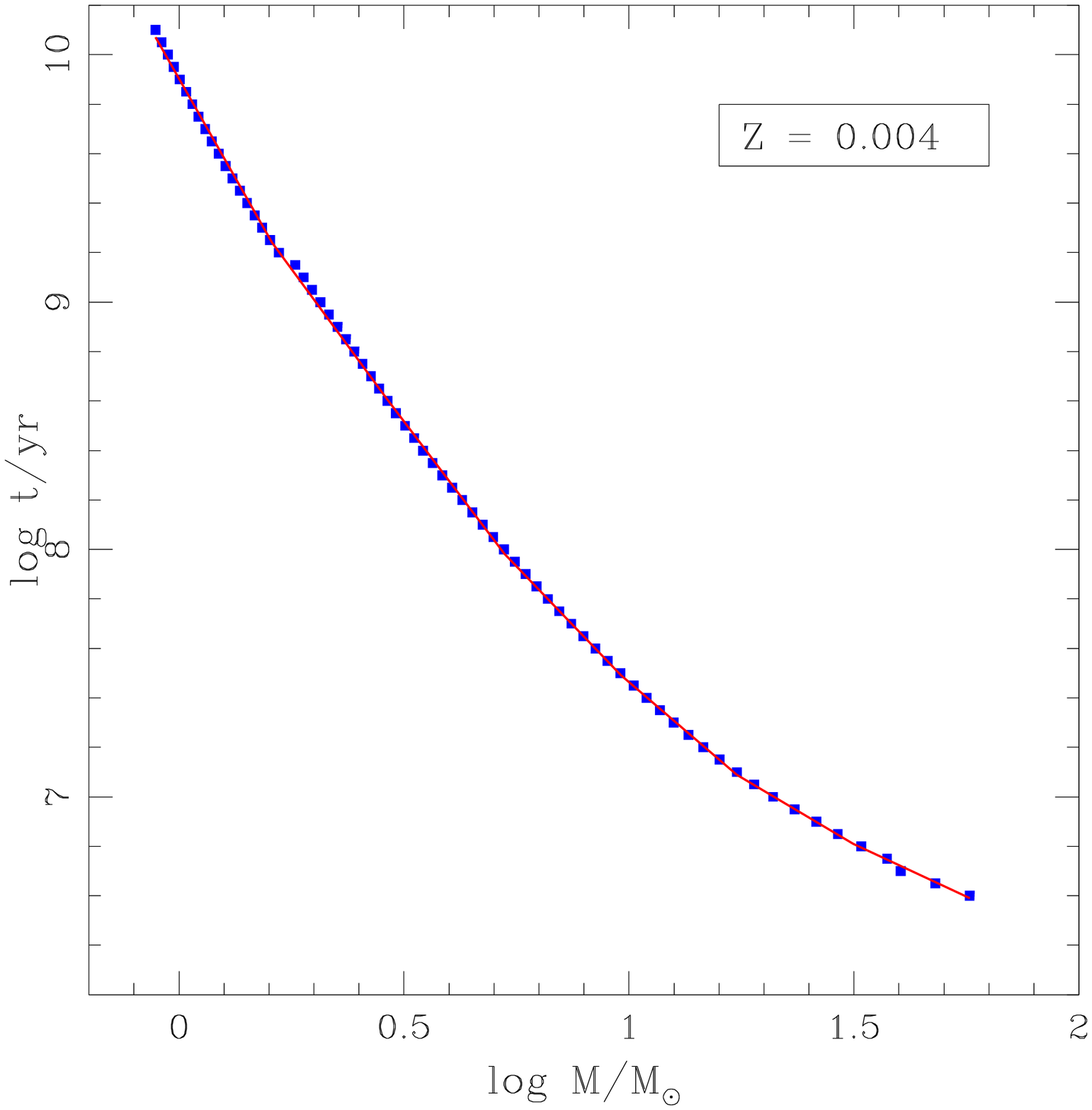,width=58mm}
\epsfig{figure=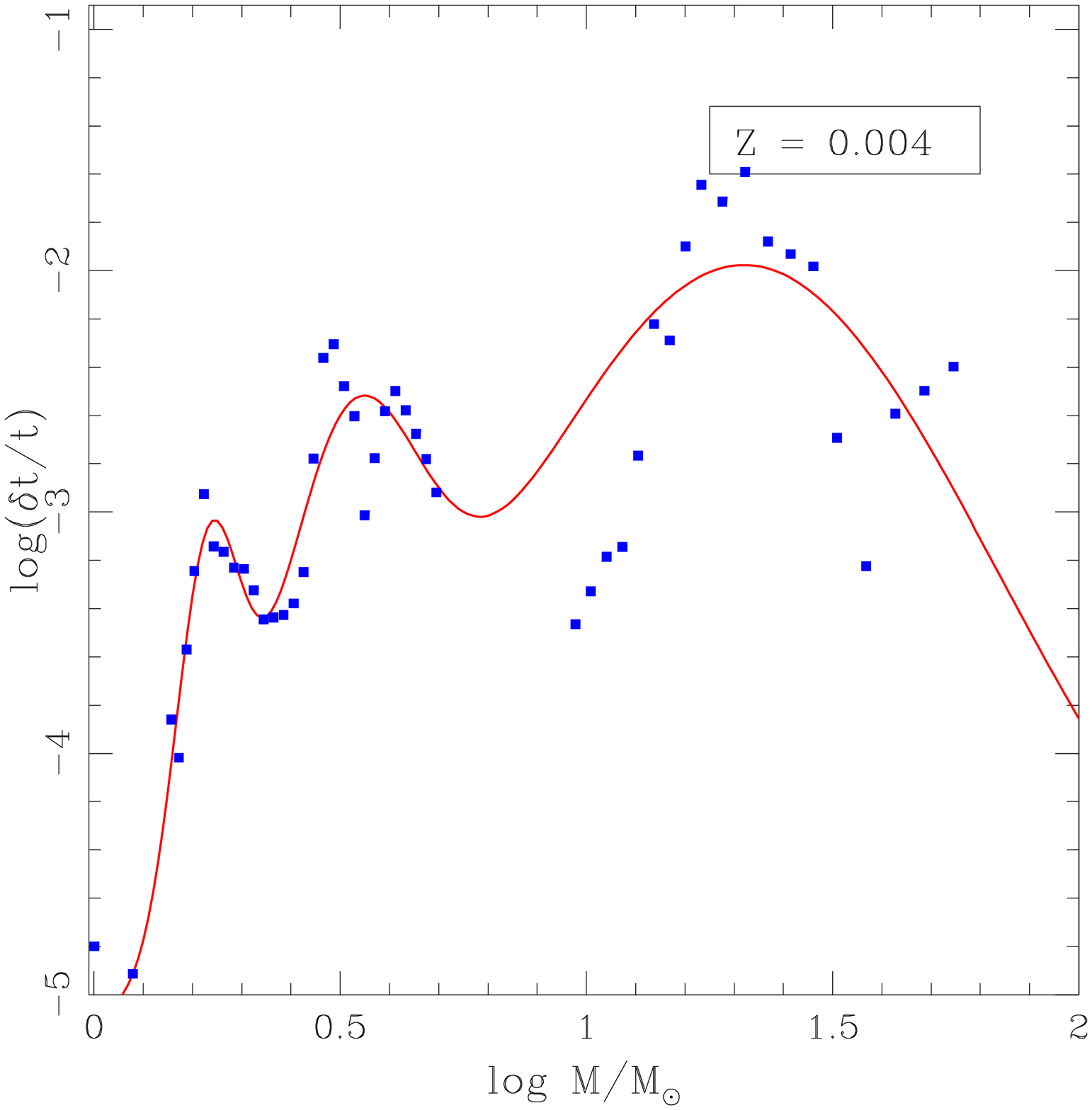,width=58mm}
}}
\caption[]{same as Figure A1 for $Z=0.004$.}
\end{figure*}

% FIGURE A5
\begin{figure*}
\centerline{\hbox{
\epsfig{figure=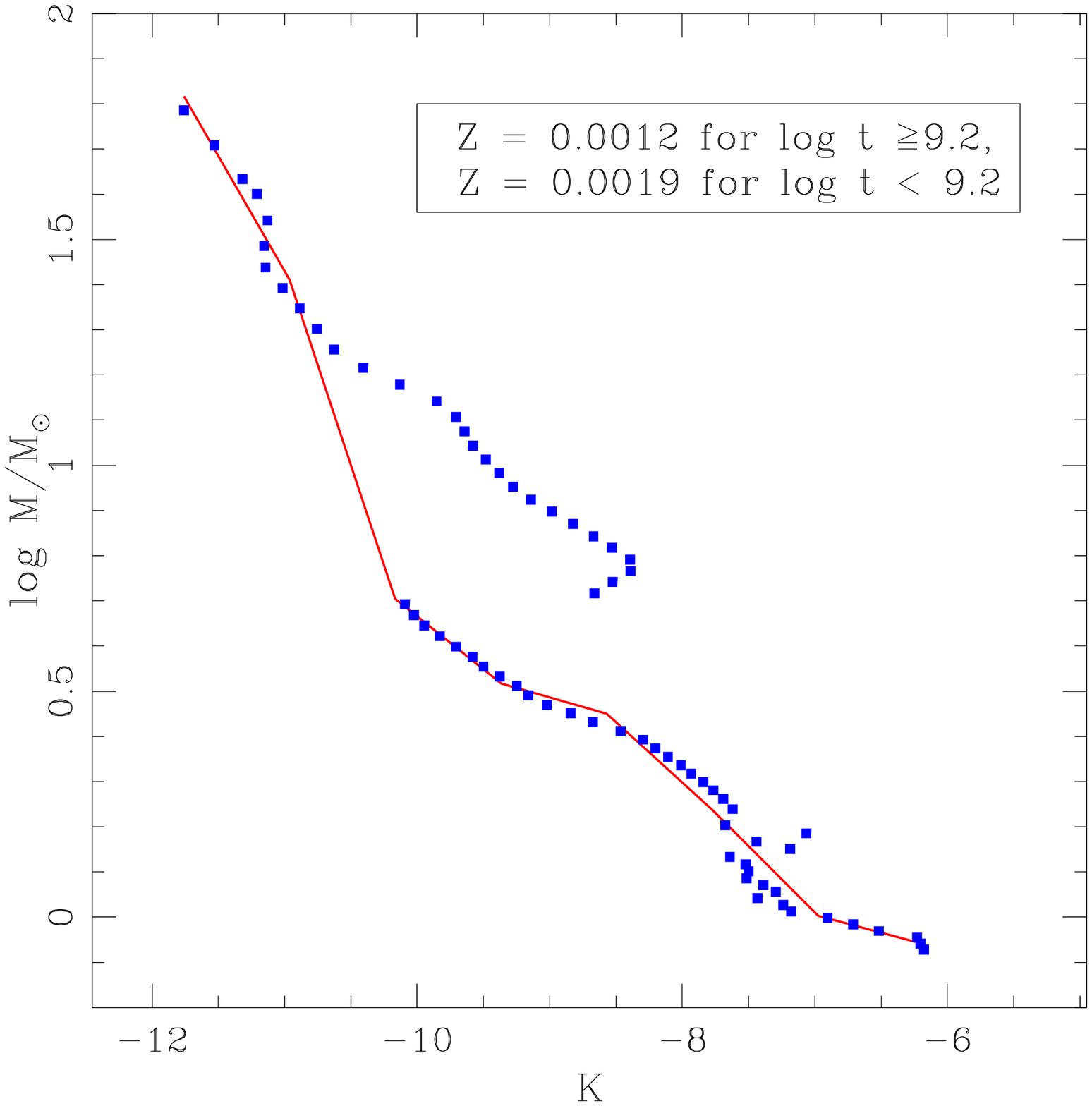,width=58mm}
\epsfig{figure=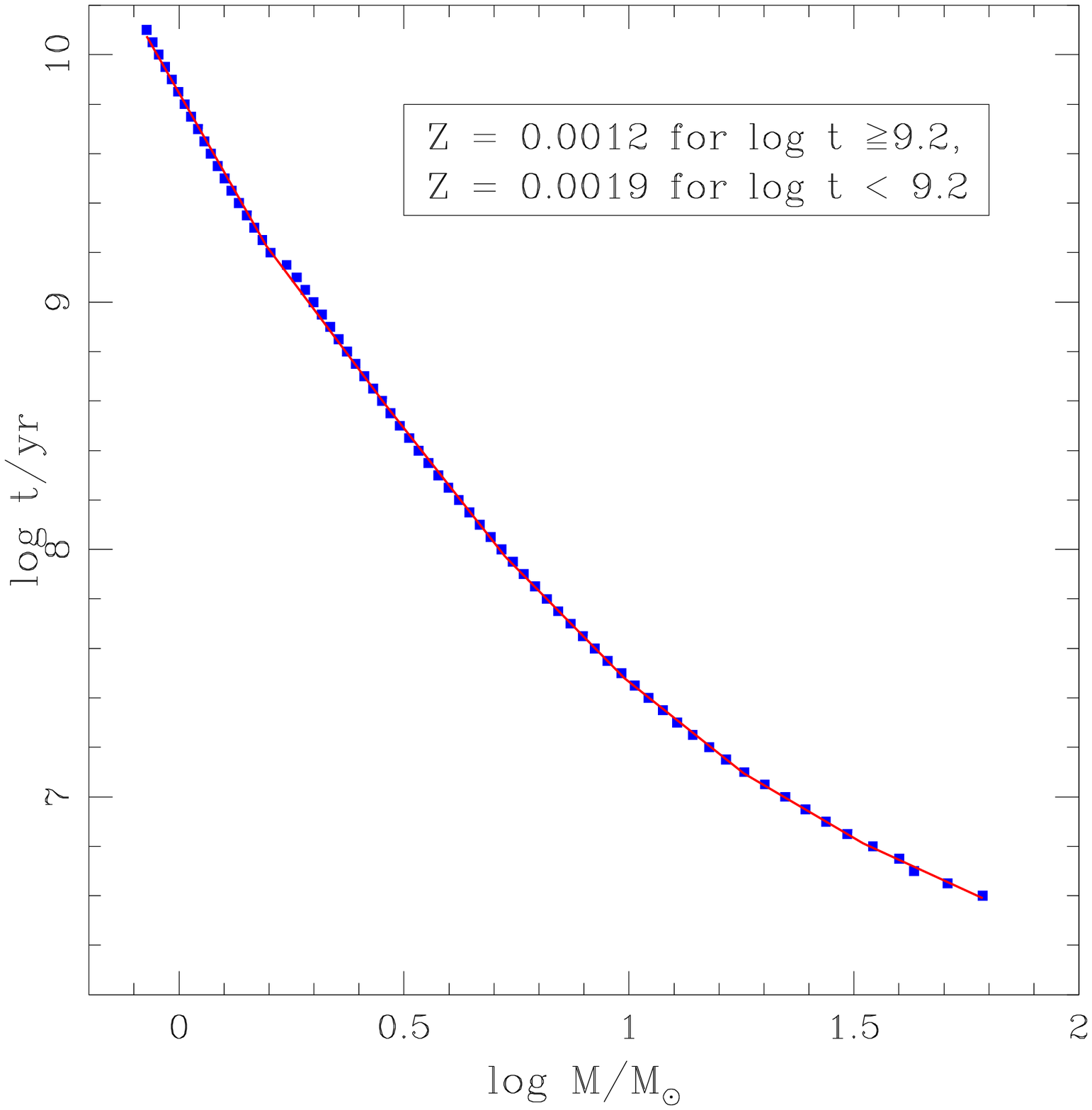,width=58mm}
\epsfig{figure=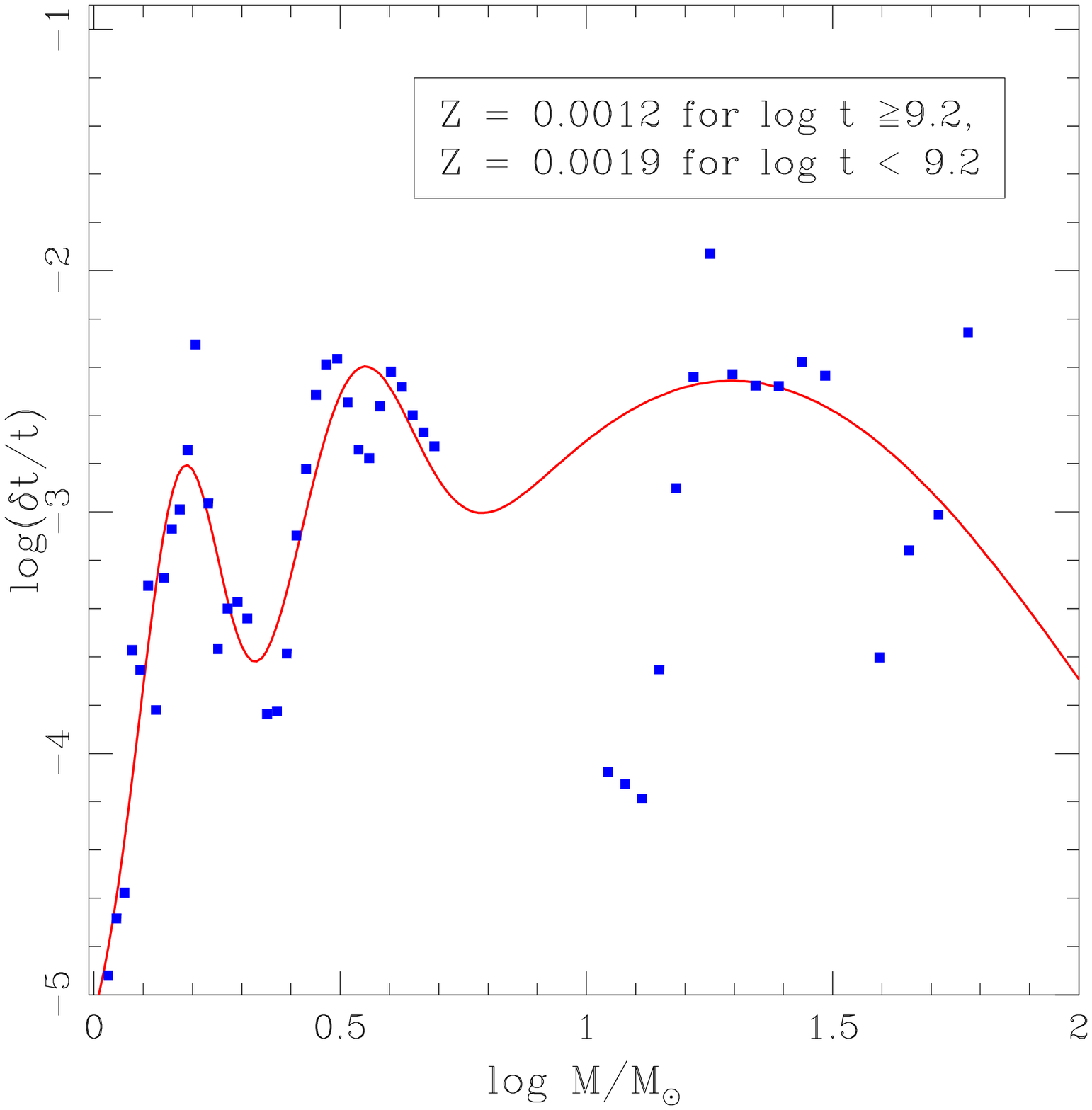,width=58mm}
}}
\caption[]{Same as Figure A1 with $Z=0.0012$ for $\log t \geq 9.2$ and
$Z=0.0019$ for $\log t < 9.2$.}
\centering
\end{figure*}

% FIGURE A6
\begin{figure*}
\centerline{\hbox{
\epsfig{figure=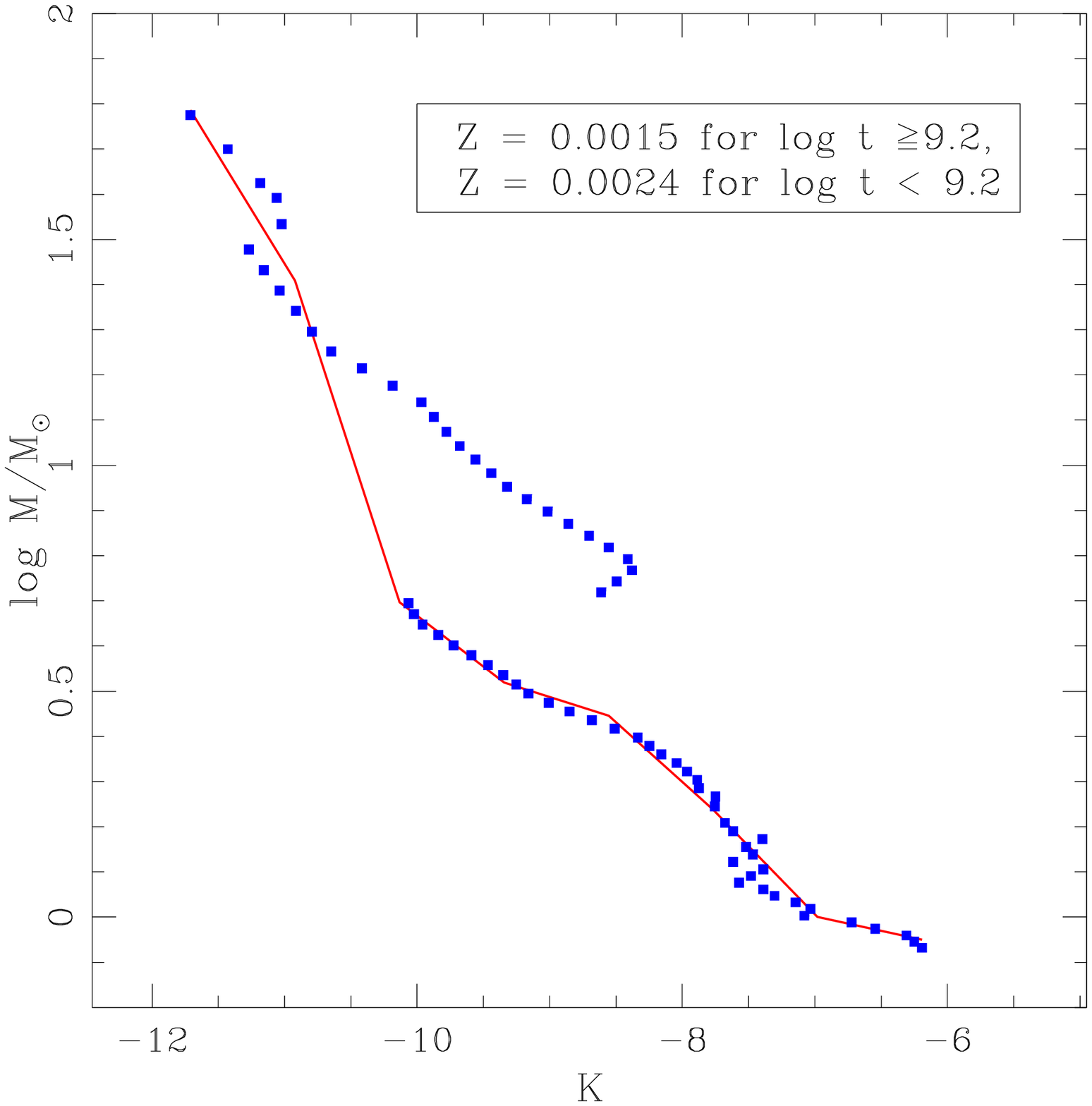,width=58mm}
\epsfig{figure=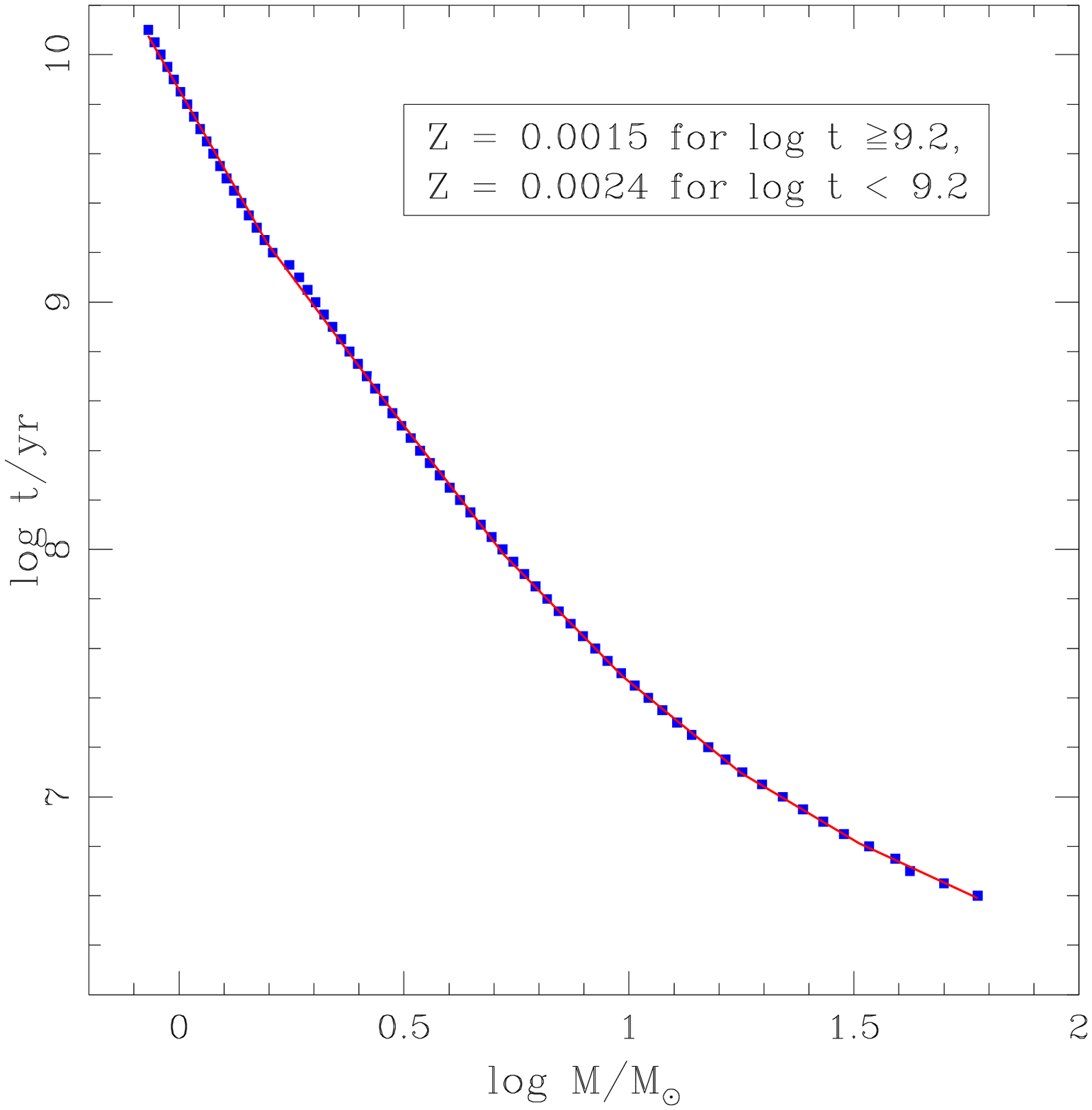,width=58mm}
\epsfig{figure=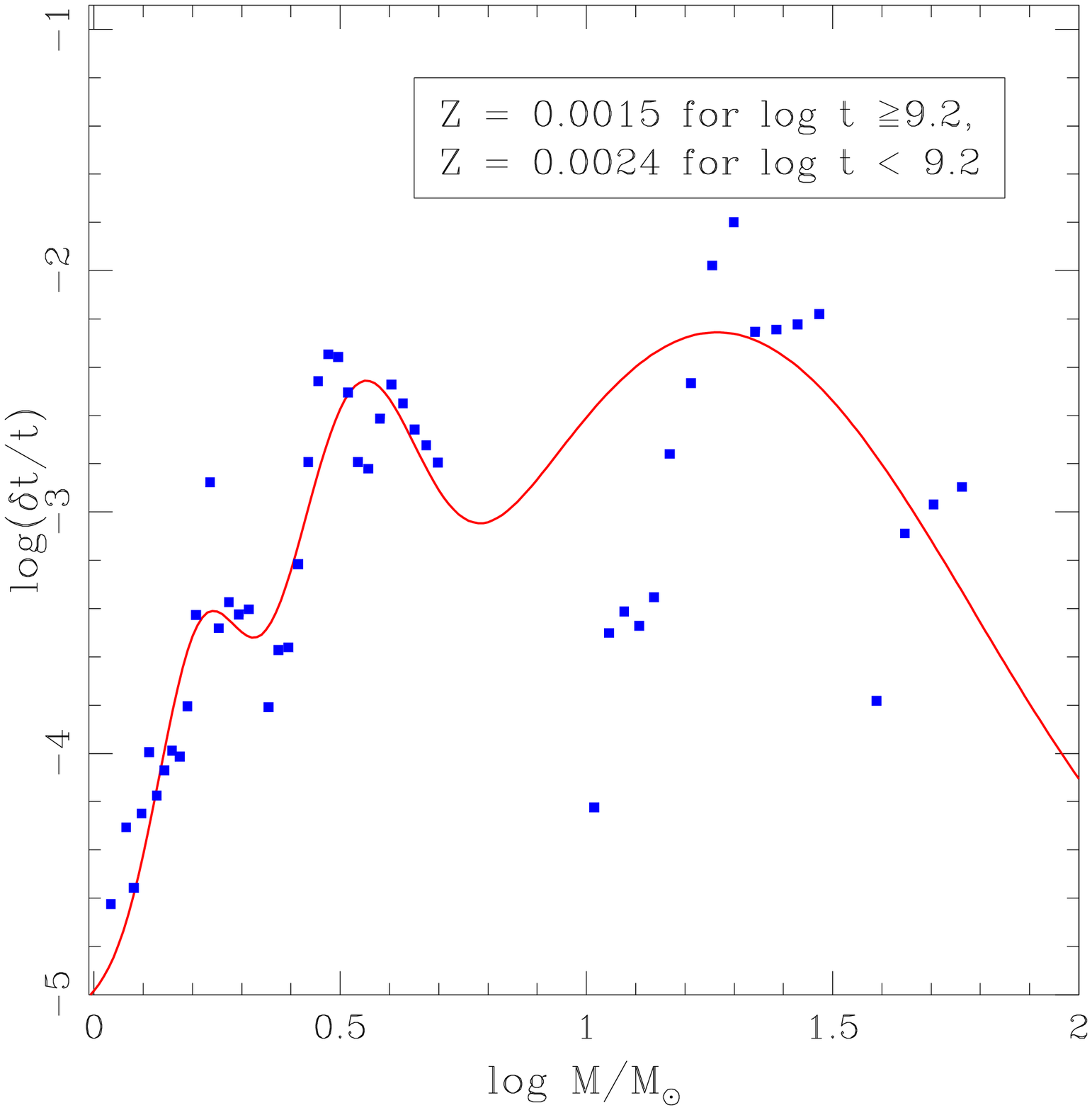,width=58mm}
}}
\caption[]{Same as Figure A1 with $Z=0.0015$ for $\log t \geq 9.2$ and
$Z=0.0024$ for $\log t < 9.2$.}
\centering
\end{figure*}

\clearpage

% TABLE A1
\begin{table}
\caption[]{Relation between birth mass and $K$-band magnitude, $\log M=aK+b$,
for a distance modulus of $\mu=0$ mag.}
\begin{tabular}{ccr}
\hline\hline
$a$              & $b$              & validity range         \\ 
\hline
\multicolumn{3}{c}{$Z=0.0012$} \\
\hline
$-0.447\pm0.086$ & $-3.502\pm0.985$ &         $K\leq-11.051$ \\
$-0.897\pm0.084$ & $-8.476\pm0.904$ & $-11.051<K\leq-10.238$ \\
$-0.233\pm0.075$ & $-1.683\pm0.737$ & $-10.238<K\leq-9.426$  \\
$-0.085\pm0.071$ & $-0.286\pm0.643$ &  $-9.426<K\leq-8.614$  \\
$-0.231\pm0.062$ & $-1.542\pm0.503$ &  $-8.614<K\leq-7.802$  \\
$-0.315\pm0.053$ & $-2.198\pm0.392$ &  $-7.802<K\leq-6.989$  \\
$-0.080\pm0.063$ & $-0.553\pm0.419$ &         $K>-6.989$     \\
\hline
\multicolumn{3}{c}{$Z=0.0015$} \\
\hline
$-0.447\pm0.071$ & $-3.477\pm0.808$ &         $K\leq-11.006$ \\
$-0.918\pm0.066$ & $-8.667\pm0.698$ & $-11.006<K\leq-10.204$ \\
$-0.232\pm0.050$ & $-1.663\pm0.486$ & $-10.204<K\leq-9.401$  \\
$-0.094\pm0.047$ & $-0.362\pm0.426$ &  $-9.401<K\leq-8.599$  \\
$-0.219\pm0.041$ & $-1.436\pm0.329$ &  $-8.599<K\leq-7.796$  \\
$-0.333\pm0.036$ & $-2.330\pm0.272$ &  $-7.796<K\leq-6.994$  \\
$-0.062\pm0.046$ & $-0.432\pm0.301$ &         $K>-6.994$     \\
\hline
\multicolumn{3}{c}{$Z=0.0024$} \\
\hline
$-0.475\pm0.070$ & $-3.777\pm0.787$ &         $K\leq-10.926$ \\
$-0.911\pm0.061$ & $-8.538\pm0.645$ & $-10.926<K\leq-10.143$ \\
$-0.225\pm0.063$ & $-1.585\pm0.612$ & $-10.143<K\leq-9.359$  \\
$-0.098\pm0.060$ & $-0.393\pm0.536$ &  $-9.359<K\leq-8.576$  \\
$-0.253\pm0.051$ & $-1.722\pm0.412$ &  $-8.576<K\leq-7.792$  \\
$-0.256\pm0.042$ & $-1.743\pm0.308$ &  $-7.792<K\leq-7.009$  \\
$-0.147\pm0.057$ & $-0.985\pm0.382$ &         $K>-7.009$     \\
\hline
\multicolumn{3}{c}{$Z=0.004$} \\
\hline
$-0.544\pm0.087$ & $-4.621\pm0.971$ &         $K\leq-10.819$ \\
$-0.803\pm0.074$ & $-7.423\pm0.771$ & $-10.819<K\leq-10.083$ \\
$-0.215\pm0.073$ & $-1.492\pm0.707$ & $-10.083<K\leq-9.347$  \\
$-0.076\pm0.070$ & $-0.194\pm0.634$ &  $-9.347<K\leq-8.610$  \\
$-0.321\pm0.061$ & $-2.304\pm0.501$ &  $-8.610<K\leq-7.874$  \\
$-0.213\pm0.052$ & $-1.452\pm0.396$ &  $-7.874<K\leq-7.138$  \\
$-0.142\pm0.071$ & $-0.949\pm0.489$ &         $K>-7.138$     \\
\hline
\multicolumn{3}{c}{$Z=0.0012$ for $\log t \geq 9.2$ and $Z=0.0019$ for
$\log t < 9.2$} \\
\hline
$-0.510\pm0.079$ & $-4.176\pm0.891$ &         $K\leq-10.962$ \\
$-0.886\pm0.071$ & $-8.298\pm0.756$ & $-10.962<K\leq-10.165$ \\
$-0.235\pm0.071$ & $-1.687\pm0.686$ & $-10.165<K\leq-9.367$  \\
$-0.084\pm0.066$ & $-0.270\pm0.599$ &  $-9.367<K\leq-8.568$  \\
$-0.267\pm0.058$ & $-1.834\pm0.467$ &  $-8.568<K\leq-7.772$  \\
$-0.294\pm0.052$ & $-2.050\pm0.385$ &  $-7.772<K\leq-6.975$  \\
$-0.079\pm0.063$ & $-0.545\pm0.416$ &         $K>-6.975$     \\
\hline
\multicolumn{3}{c}{$Z=0.0015$ for $\log t \geq 9.2$ and $Z=0.0024$ for
$\log t < 9.2$} \\
\hline
$-0.478\pm0.073$ & $-3.810\pm0.827$ &         $K\leq-10.922$ \\
$-0.903\pm0.063$ & $-8.458\pm0.669$ & $-10.922<K\leq-10.133$ \\
$-0.225\pm0.065$ & $-1.579\pm0.633$ & $-10.133<K\leq-9.345$  \\
$-0.095\pm0.062$ & $-0.365\pm0.556$ &  $-9.345<K\leq-8.556$  \\
$-0.262\pm0.053$ & $-1.795\pm0.429$ &  $-8.556<K\leq-7.768$  \\
$-0.303\pm0.051$ & $-2.114\pm0.381$ &  $-7.768<K\leq-6.979$  \\
$-0.064\pm0.064$ & $-0.444\pm0.425$ &         $K>-6.979$     \\
\hline
\end{tabular}
\end{table}

% TABLE A2
\begin{table}
\caption[]{Relation between age and birth mass, $\log t=a\log M+b$.}
\begin{tabular}{ccr}
\hline \hline
$a$              & $b$             & validity range           \\
\hline
\multicolumn{3}{c}{$Z=0.0012$} \\
\hline
$-3.109\pm0.033$ & $9.842\pm0.003$ &       $\log{M}\leq0.241$ \\
$-2.433\pm0.032$ & $9.678\pm0.013$ & $0.241<\log{M}\leq0.555$ \\
$-2.043\pm0.034$ & $9.462\pm0.024$ & $0.555<\log{M}\leq0.868$ \\
$-1.618\pm0.039$ & $9.093\pm0.040$ & $0.868<\log{M}\leq1.181$ \\
$-1.074\pm0.046$ & $8.451\pm0.061$ & $1.181<\log{M}\leq1.495$ \\
$-0.833\pm0.060$ & $8.090\pm0.097$ &       $\log{M}>1.495$    \\
\hline
\multicolumn{3}{c}{$Z=0.0015$} \\
\hline
$-3.216\pm0.030$ & $9.860\pm0.002$ &       $\log{M}\leq0.198$ \\
$-2.511\pm0.028$ & $9.720\pm0.009$ & $0.198<\log{M}\leq0.465$ \\
$-2.238\pm0.030$ & $9.593\pm0.018$ & $0.465<\log{M}\leq0.732$ \\
$-1.850\pm0.033$ & $9.310\pm0.028$ & $0.732<\log{M}\leq0.998$ \\
$-1.413\pm0.038$ & $8.873\pm0.042$ & $0.998<\log{M}\leq1.265$ \\
$-1.063\pm0.043$ & $8.430\pm0.060$ & $1.265<\log{M}\leq1.531$ \\
$-0.801\pm0.054$ & $8.030\pm0.088$ &       $\log{M}>1.531$    \\
\hline
\multicolumn{3}{c}{$Z=0.0024$} \\
\hline
$-3.100\pm0.035$ & $9.881\pm0.004$ &       $\log{M}\leq0.245$ \\
$-2.479\pm0.033$ & $9.728\pm0.013$ & $0.245<\log{M}\leq0.551$ \\
$-2.128\pm0.035$ & $9.535\pm0.025$ & $0.551<\log{M}\leq0.857$ \\
$-1.649\pm0.040$ & $9.124\pm0.040$ & $0.857<\log{M}\leq1.163$ \\
$-1.156\pm0.047$ & $8.552\pm0.061$ & $1.163<\log{M}\leq1.469$ \\
$-0.862\pm0.061$ & $8.119\pm0.097$ &       $\log{M}>1.469$    \\
\hline
\multicolumn{3}{c}{$Z=0.004$} \\
\hline
$-3.207\pm0.045$ & $9.904\pm0.004$ &       $\log{M}\leq0.207$ \\
$-2.470\pm0.041$ & $9.751\pm0.014$ & $0.207<\log{M}\leq0.465$ \\
$-2.398\pm0.043$ & $9.718\pm0.025$ & $0.465<\log{M}\leq0.723$ \\
$-1.901\pm0.047$ & $9.358\pm0.040$ & $0.723<\log{M}\leq0.982$ \\
$-1.559\pm0.053$ & $9.023\pm0.059$ & $0.982<\log{M}\leq1.240$ \\
$-1.083\pm0.062$ & $8.432\pm0.084$ & $1.240<\log{M}\leq1.499$ \\
$-0.846\pm0.078$ & $8.076\pm0.126$ &       $\log{M}>1.499$    \\
\hline
\multicolumn{3}{c}{$Z=0.0012$ for $\log t \geq 9.2$ and $Z=0.0019$ for
$\log t < 9.2$} \\
\hline
$-3.180\pm0.041$ & $9.845\pm0.003$ &       $\log{M}\leq0.193$ \\
$-2.428\pm0.038$ & $9.699\pm0.012$ & $0.193<\log{M}\leq0.459$ \\
$-2.312\pm0.039$ & $9.646\pm0.023$ & $0.459<\log{M}\leq0.724$ \\
$-1.854\pm0.044$ & $9.314\pm0.037$ & $0.724<\log{M}\leq0.990$ \\
$-1.451\pm0.050$ & $8.915\pm0.055$ & $0.990<\log{M}\leq1.255$ \\
$-1.063\pm0.057$ & $8.429\pm0.079$ & $1.255<\log{M}\leq1.521$ \\
$-0.834\pm0.072$ & $8.080\pm0.117$ &       $\log{M}>1.521$    \\
\hline
\multicolumn{3}{c}{$Z=0.0015$ for $\log t \geq 9.2$ and $Z=0.0024$ for
$\log t < 9.2$} \\
\hline
$-3.183\pm0.040$ & $9.860\pm0.003$ &       $\log{M}\leq0.195$ \\
$-2.437\pm0.037$ & $9.715\pm0.012$ & $0.195<\log{M}\leq0.459$ \\
$-2.345\pm0.039$ & $9.672\pm0.022$ & $0.459<\log{M}\leq0.722$ \\
$-1.868\pm0.043$ & $9.328\pm0.036$ & $0.722<\log{M}\leq0.985$ \\
$-1.478\pm0.049$ & $8.944\pm0.054$ & $0.985<\log{M}\leq1.249$ \\
$-1.082\pm0.056$ & $8.448\pm0.077$ & $1.249<\log{M}\leq1.512$ \\
$-0.842\pm0.070$ & $8.086\pm0.114$ &       $\log{M}>1.512$    \\
\hline
\end{tabular}
\end{table}

% TABLE A3
\begin{table}
\caption[]{Relation between the relative pulsation duration and birth mass,
$\log(\delta t/t)=D+
\Sigma_{i=1}^3a _i\exp\left[-(\log M [{\rm M} _\odot]-b_i)^2/2c_i^2\right]$.}
\begin{tabular}{ccccc}
\hline\hline
$D$    & $i$ & $a$   & $b$   & $c$   \\
\hline
\multicolumn{5}{c}{$Z=0.0012$} \\
\hline
$-5.2$ &  1  & 1.955 & 0.173 & 0.074 \\
       &  2  & 2.545 & 0.545 & 0.196 \\
       &  3  & 2.936 & 1.539 & 0.429 \\
\hline
\multicolumn{5}{c}{$Z=0.0015$} \\
\hline
$-4.9$ &  1  & 0.850 & 0.202 & 0.080 \\
       &  2  & 2.265 & 0.547 & 0.188 \\
       &  3  & 2.748 & 1.483 & 0.410 \\
\hline
\multicolumn{5}{c}{$Z=0.0024$} \\
\hline
$-4.8$ &  1  & 1.267 & 0.222 & 0.064 \\
       &  2  & 1.831 & 0.529 & 0.147 \\
       &  3  & 2.567 & 1.302 & 0.427 \\
\hline
\multicolumn{5}{c}{$Z=0.004$} \\
\hline
$-5.2$ &  1  & 1.510 & 0.229 & 0.065 \\
       &  2  & 1.702 & 0.508 & 0.149 \\
       &  3  & 3.258 & 1.320 & 0.519 \\
\hline
\multicolumn{5}{c}{$Z=0.0012$ for $\log t \geq 9.2$ and $Z=0.0019$ for
$\log t < 9.2$} \\
\hline
$-6.5$ &  1  & 2.020 & 0.175 & 0.083 \\
       &  2  & 1.455 & 0.519 & 0.126 \\
       &  3  & 4.011 & 1.295 & 0.816 \\
\hline
\multicolumn{5}{c}{$Z=0.0015$ for $\log t \geq 9.2$ and $Z=0.0024$ for
$\log t < 9.2$} \\
\hline
$-5.2$ &  1  & 1.240 & 0.213 & 0.086 \\
       &  2  & 1.631 & 0.520 & 0.133 \\
       &  3  & 2.944 & 1.265 & 0.522 \\
\hline
\end{tabular}
\end{table}

\label{lastpage}
\end{document}